\documentclass[12pt,english]{article}
\usepackage[pdftex]{geometry}
\usepackage{bm}
\usepackage{amsmath}
\usepackage{amssymb}
\usepackage[pdftex]{graphicx}
\usepackage{setspace}
\makeatletter
\numberwithin{equation}{section}
\newcommand{\lyxaddress}[1]{
	\par {\raggedright #1
	\vspace{1.4em}
	\noindent\par}
}

\allowdisplaybreaks[1]

\usepackage{color}
\usepackage[pdftex]{hyperref}
\hypersetup{
colorlinks,
linktoc=page,
citecolor=blue,
linkcolor=blue,
urlcolor=blue
}

\date{}

\makeatother

\usepackage{babel}
\usepackage[style=phys,biblabel=brackets, chaptertitle=false, pageranges=false, eprint=true]{biblatex}
\begin{document}
\begin{flushright}
{\small{}YITP-20-87}{\small\par}
\par\end{flushright}

\noindent\begin{minipage}[t]{1\columnwidth}%
\title{\textbf{Fermi gas approach to general rank theories and quantum curves}}
\author{Naotaka Kubo\,\footnotemark}
\maketitle

\lyxaddress{\begin{center}
\vspace{-18bp}
$^{*}$\,\textit{Center for Gravitational Physics, Yukawa Institute for Theoretical Physics,}\\
\textit{Kyoto University, Sakyo-ku, Kyoto 606-8502, Japan}\vspace{-10bp}
\par\end{center}}
\begin{abstract}
It is known that matrix models computing the partition functions of three-dimensional $\mathcal{N}=4$ superconformal Chern-Simons theories described by circular quiver diagrams can be written as the partition functions of ideal Fermi gases when all the nodes have equal ranks. We extend this approach to rank deformed theories. The resulting matrix models factorize into factors depending only on the relative ranks in addition to the Fermi gas factors. We find that this factorization plays a critical role in showing the equality of the partition functions of dual theories related by the Hanany-Witten transition. Furthermore, we show that the inverses of the density matrices of the ideal Fermi gases can be simplified and regarded as quantum curves as in the case without rank deformations. We also comment on four nodes theories using our results.
\end{abstract}
\end{minipage}

\renewcommand{\thefootnote}{\fnsymbol{footnote}}
\footnotetext[1]{\textsf{naotaka.kubo@yukawa.kyoto-u.ac.jp}}
\renewcommand{\thefootnote}{\arabic{footnote}}

\newpage{}

\global\long\def\ABJM{\mathrm{ABJ}}%

\global\long\def\Iside{\mathrm{vec}}%

\global\long\def\Icross{\mathrm{mat}}%

\global\long\def\bra#1{\langle#1|}%

\global\long\def\bbra#1{\langle\negthinspace\langle#1|}%

\global\long\def\ket#1{|#1\rangle}%

\global\long\def\kket#1{|#1\rangle\negthinspace\rangle}%

\global\long\def\braket#1{\langle#1\rangle}%

\global\long\def\bbraket#1{\langle\negthinspace\langle#1\rangle}%

\global\long\def\brakket#1{\langle#1\rangle\negthinspace\rangle}%

\global\long\def\bbrakket#1{\langle\negthinspace\langle#1\rangle\negthinspace\rangle}%

\tableofcontents{}

\section{Introduction}

M2-branes are important objects in M-theory, and thus it is indispensable for studying M-theory to understand their non-perturbative properties. One of the most important progress in the study of M2-branes is the discovery of its worldvolume theory. M2-branes on $\mathbb{C}^{4}/\mathbb{Z}_{k}$ can be obtained by taking T-duality and the M-theory lift to a brane configuration consisting of D3-branes with one direction compactified, an NS5-brane and a $\left(1,k\right)$5-brane in type IIB string theory. The worldvolume theory of this brane configuration is the $\mathcal{N}=6$ superconformal Chern-Simons theory with gauge group $\mathrm{U}\left(N\right)_{k}\times\mathrm{U}\left(N\right)_{-k}$ called ABJM theory \cite{Aharony:2008ug}. The subscriptions of the unitary groups denote their Chern-Simons levels.

The partition function of the ABJM theory on $S^{3}$ reduces to the ABJM matrix model using the localization technique \cite{Kapustin:2009kz}. Analysis in the 't Hooft limit
\begin{equation}
N\rightarrow\infty,\quad\lambda=\frac{N}{k}:\mathrm{fixed},
\end{equation}
which is the traditional way to study matrix models, however, is not suitable, because the gravitational dual of the ABJM theory is M-theory on $AdS_{4}\times S^{7}/\mathbb{Z}_{k}$ and thus M-theory reduces to type IIA string theory in $AdS_{4}\times\mathbb{CP}^{3}$ with the $k\rightarrow\infty$ limit.

A new approach called Fermi gas formalism was developed in \cite{Marino:2011eh} to resolve this problem. They suggested to rewrite the ABJM matrix model as the partition function of an ideal Fermi gas with a non-trivial one-particle density matrix
\begin{equation}
\widehat{\rho}^{\mathrm{ABJM}}=\frac{1}{2\cosh\frac{\widehat{p}}{2}}\frac{1}{2\cosh\frac{\widehat{q}}{2}},\label{eq:rhoABJM}
\end{equation}
with Plank constant $\hbar=2\pi k$. After this reformulation, we can analyze the small $\hbar$ regime, in other words, the small $k$ regime in the WKB or semiclassical expansion because this system is now the quantum mechanical system. This approach enables direct derivation of the $N^{\frac{3}{2}}$ behavior of the free energy \cite{Marino:2011eh}. It was found that the Fermi gas formalism was the powerful tool also for calculating the exact values of the partition functions for various $k$ and $N$ \cite{Hatsuda:2012hm,Putrov:2012zi,Hatsuda:2012dt}, membrane instantons \cite{Marino:2011eh,Hatsuda:2012dt,Calvo:2012du}, the vacuum expectation values of Wilson loops \cite{Klemm:2012ii,Hatsuda:2013yua,Kiyoshige:2016lno,Kubo:2018gdn} and other physical quantities \cite{Chester:2020jay} (see also \cite{Hatsuda:2015gca,Marino:2016new} for reviews). All these successes demonstrate that the Fermi gas formalism is an exceedingly useful approach to study the M-theory regime.

There are many generalizations of the ABJM theory. In this paper, we consider two steps of generalizations. The first generalization is to deform the rank of one node, which is called  ABJ theory \cite{Hosomichi:2008jb,Aharony:2008gk}. It was conjectured that the ABJ theories with gauge groups $\mathrm{U}\left(N\right)_{k}\times\mathrm{U}\left(N+M\right)_{-k}$ and $\mathrm{U}\left(N+k-M\right)_{k}\times\mathrm{U}\left(N\right)_{-k}$ were Seiberg-like dual theories \cite{Giveon:2008zn}, which is the consequence of the Hanany-Witten transition \cite{Hanany:1996ie} in type IIB string theory. To confirm this conjecture, we should deal with rank deformed theories. However, it was more difficult to apply the Fermi gas formalism to the ABJ matrix model than the case of the ABJM matrix model. Nevertheless, in \cite{Awata:2012jb,Honda:2013pea,Honda:2014npa},\footnote{Another formalism was developed in \cite{Matsumoto:2013nya}.} they succeeded to rewrite the ABJ matrix model associated with the gauge group $\mathrm{U}\left(N\right)_{k}\times\mathrm{U}\left(N+M\right)_{-k}$ as the partition function of an ideal Fermi gas with the density matrix
\begin{equation}
\widehat{\rho}_{M}^{\mathrm{ABJ}}=i^{M}\frac{1}{\prod_{j=1}^{M}2\cosh\frac{\widehat{q}-2\pi it_{M,j}}{2k}}\frac{1}{2\cosh\frac{\widehat{p}}{2}}\frac{\prod_{j=1}^{M}2\sinh\frac{\widehat{q}-2\pi it_{M,j}}{2k}}{2\cosh\frac{\widehat{q}+i\pi M}{2}},\label{eq:rhoABJ}
\end{equation}
where $t_{M,j}=\frac{1}{2}\left(M+1\right)-j$. They also indicated that the density matrices of the dual theories were identical. This equality leads to the equality of the partition functions of the dual theories (up to a phase factor).

The inverse of the ABJM density matrix \eqref{eq:rhoABJM} is the Laurent polynomial of $e^{\frac{1}{2}\widehat{q}}$ and $e^{\frac{1}{2}\widehat{p}}$. Laurent polynomials of these operators are called quantum curves. On the other hand, it is not obvious that the inverse of the ABJ density matrix \eqref{eq:rhoABJ} can also be regarded as a quantum curve. Nevertheless, it was found in \cite{Kashaev:2015wia} that the inverse matrix also can be put in the form of a quantum curve. This progress allowed us to discuss, for example, relations between the ABJ theory and the TS/ST correspondence \cite{Grassi:2014zfa} and between the ABJ theory and the $q$-deformed Painlev\'{e} equations \cite{Bonelli:2017gdk}.

As we explained above, the ABJ(M) theory is the worldvolume theory of the system containing D3-branes with one direction compactified, a single NS5-brane and a single $\left(1,k\right)$5-brane. The second generalization is to increase the number of 5-branes arbitrarily \cite{Imamura:2008dt}. The worldvolume theories of these brane configurations are $\mathcal{N}=4$ superconformal Chern-Simons theories described by circular quiver diagrams. The matrix models associated with these gauge theories also can be written as the partition functions of ideal Fermi gases when all the nodes have equal ranks, and much studied in the Fermi gas description \cite{Marino:2011eh,Grassi:2014vwa,Honda:2014ica,Hatsuda:2014vsa,Moriyama:2014gxa,Moriyama:2014nca,Hatsuda:2015lpa} as with the ABJ(M) theory. However, when the ranks are different, so far the Fermi gas formalism is applied only to special cases such as theories with gauge groups consisting of four unitary groups \cite{Moriyama:2017gye,Kubo:2019ejc}.

On the other hand, there was recently progress in the quantum curves. It was revealed that some types of quantum curves obey the Weyl group symmetry of $\mathrm{SO}\left(10\right)$ and the exceptional groups \cite{Kubo:2018cqw,Moriyama:2020lyk}. Furthermore, it was found in \cite{Kubo:2019ejc} by connecting the $\mathrm{SO}\left(10\right)$ quantum curve to the rank deformed theories with four nodes without using the Fermi gas formalism that the Hanany-Witten transition is realized as a portion of the Weyl group symmetry of $\mathrm{SO}\left(10\right)$. This fact implies that quantum curves play an important role in understanding M2-brane physics.

Given the above considerations, we expect that the Fermi gas description and quantum curves have sufficient potential to study non-perturbative effects such as the Hanany-Witten transition. Therefore, it is important to apply the Fermi gas formalism not only to theories without any rank deformations but also to rank deformed theories and put the density matrices in the form of quantum curves. In this paper, we realize these ideas. 

The strategy we use in this paper to apply the Fermi gas formalism is to cut the integrand of the matrix model at positions where corresponding ranks of nodes are lowest and locally apply the Fermi gas formalism to each part of matrix model, which we call deformed factor (because each deformed factor corresponds to each lump of 5-brane factors with ranks deformed from the lowest rank). Looking at this procedure from a different angle, we can apply the Fermi gas formalism to all of the rank deformations constructed by the deformed factors. Therefore, this procedure can be applied to a wide class of rank deformations. This procedure first appeared in \cite{Kubo:2019ejc} though they computed the deformed factor associated with the brane configuration consisting of only one NS5-brane and one $\left(1,k\right)$5-brane. In this paper, we focus on brane configurations consisting of an NS5-brane at the center, any number of $\left(1,k\right)$5-branes on both side of the NS5-brane and D3-branes stretched between two 5-branes such that the number monotonically increases with approaching the NS5-brane (see figure \ref{fig:BraneConfig} in section \ref{subsec:Target}). The corresponding deformed factors are \eqref{eq:ZpartGen}.

Our main result is that, after similarity transformations which we explain in section \ref{subsec:Strategy}, \eqref{eq:ZpartGen} is transformed into \eqref{eq:ZpGenRes}, which is nothing but the Fermi gas description. The non-trivial factorization appears in this expression, and this factorization displays its potential when we study dualities. We find that we can divide the factors into three types of factors, namely $Z_{k,M}^{\left(\mathrm{CS}\right)}$, $Z_{k,M,M'}^{\left(\Iside/\Icross\right)}$ and the determinant of the density matrix, and they obey identities expected from the Hanany-Witten transition separately. In other words, the equality between dual theories holds for each type of factor. We also comment on the good/ugly/bad classification given in \cite{Gaiotto:2008ak} by studying divergences appearing in these factors.

The other main result is that the inverses of the density matrices \eqref{eq:rhoGen} can be regarded as quantum curves as in \eqref{eq:QCgen}. Comparing these quantum curves to the brane configurations, their relation can be visually understood. By using this result, we study specific $\mathcal{N}=4$ Chern-Simons theories with four unitary groups. First, we partially prove the relation between these gauge theories and quantum curves. Second, we find out that a Weyl group symmetry obeyed by the quantum curves strongly depends on the form of the density matrices by revising in terms of the density matrices.

This paper is organized as follows. In section \ref{sec:SCSMM}, first we briefly review brane configurations in type IIB string theory and matrix models which the partition functions of the worldvolume theories of these brane configurations reduce to using the localization technique. Second, we explain our strategy to apply the Fermi gas formalism. Third, we review the Fermi gas formalism to simple examples. In section \ref{sec:FGF}, we present new deformed factors we study in this paper and apply the Fermi gas formalism. We also prove the equality between the deformed factors related by the Hanany-Witten transition. In section \ref{sec:QC}, we put the inverses of the density matrices we obtained in the previous section in the form of quantum curves. We comment on them from the point of view of the brane configurations and argue the Hanany-Witten transition. In section \ref{sec:D5Model}, we study specific $\mathcal{N}=4$ Chern-Simons theories with four nodes. Our results provide new insights into these gauge theories. Finally, we conclude and discuss some future directions in section \ref{sec:Conclusion}. Appendix \ref{sec:Formulas} contains various formulas and their proofs. Appendix \ref{sec:integral} comments on delta functions whose argument is a complex number. Appendix \ref{sec:QCproof} provides proof for the relation between the density matrices and the quantum curves.

\section{Matrix models and Fermi gas formalism\label{sec:SCSMM}}

In this section we review the Fermi gas formalism. First, we present brane configurations in type IIB string theory which we deal with in this paper. Second, we briefly review matrix models computing the partition functions of the worldvolume theories of the brane configurations. Third, we explain our strategy to apply the Fermi gas formalism to these matrix models. Finally, we review the way we rewrite the matrix models as the partition functions of ideal Fermi gases for simple examples including the ABJ theory.

In this paper we use the following shorthand notations
\begin{equation}
\prod_{n}^{N}\ldots=\prod_{n=1}^{N}\ldots,\quad\prod_{n<n'}^{N}\ldots=\prod_{n=1}^{N-1}\prod_{n'=n+1}^{N}\ldots,
\end{equation}
and similar notations for the sums. In addition, $\left[f_{a,b}\right]_{a,b}^{A\times B}$ denotes an $A\times B$ matrix whose $\left(a,b\right)$ element is $f_{a,b}$.

\subsection{From brane configurations to matrix models}

In this section we present brane configurations studied in \cite{Imamura:2008dt} and corresponding matrix models. We also confirm our notations.

We start with the brane configurations consisting of NS5-branes along the directions $012345$ and $\left(1,k\right)$5-branes along the directions $012\left[3,7\right]_{\theta}\left[4,8\right]_{\theta}\left[5,9\right]_{\theta}$ where $\theta$ is chosen such that the $\mathcal{N}=4$ supersymmetry is preserved. The 5-branes are separated along the compact direction $x^{6}$. We assign $+1$ to an NS5-brane and $-1$ to a $\left(1,k\right)$5-brane, so that the clockwise order of 5-branes is translated into the digit sequence $\bm{s}=\left(s_{1},s_{2},\ldots,s_{r}\right)$ of $s_{a}=\pm1$ with the cyclic identification $s_{0}=s_{r}$ where $r$ denotes the number of 5-branes. Next, we insert a stack of coincident D3-branes along the directions $0126$ in each interval of 5-branes.\footnote{These brane configurations are also related to M2-branes on $\left(\mathbb{C}^{2}/\mathbb{Z}_{p}\times\mathbb{C}^{2}/\mathbb{Z}_{q}\right)/\mathbb{Z}_{k}$ via T-duality and the M-theory lift \cite{Imamura:2008ji}.} We describe the number of D3-branes for each interval in two ways. The first way is simply to enumerate the number of D3-branes for each intervals as $\bm{N}=\left(N_{1},N_{2},\ldots,N_{r}\right)$ where $N_{a}$ is the number between $s_{a-1}$ and $s_{a}$. The second way is to determine a reference number $N$ and enumerate the relative number as $\bm{M}=\left(M_{1},M_{2},\ldots,M_{r}\right)$. In this paper, the reference number $N$ is always minimum except for section \ref{sec:D5Model}. That is, the relation between the first and the second notation is
\begin{equation}
N=\min_{a}\left\{ N_{a}\right\} ,\quad M_{a}=N_{a}-N.\label{eq:LabelRelation}
\end{equation}
In summary, the brane configurations are labeled by $\bm{s}$ and $\bm{N}$, or $\bm{s}$, $N$ and $\bm{M}$. We use these two types of descriptions interchangeably. We also prepare visual expression by replacing $s_{a}=+1$ and $s_{a}=-1$ with $\bullet$ and $\circ$ respectively (in other words, $\bullet$ and $\circ$ denote an NS5-brane and a $\left(1,k\right)$5-brane respectively) and arranging the numbers of D3-branes. For example, 
\begin{equation}
\left\langle N_{1}\bullet N_{2}\circ N_{3}\circ\right\rangle ^{\mathrm{P}},
\end{equation}
means $\bm{s}=\left(+1,-1,-1\right)$ and $\bm{N}=\left(N_{1},N_{2},N_{3}\right)$. The corresponding relative description is
\begin{equation}
\left\langle M_{1}\bullet M_{2}\circ M_{3}\circ\right\rangle _{N}^{\mathrm{P}}.
\end{equation}
$\mathrm{P}$ of superscription emphasizes that $x^{6}$ direction is periodic. On the other hand, when we focus on a divided part, we describe only that part and exclude $\mathrm{P}$. We also write $\star$ instead of $\bullet$ and $\circ$ when we avoid to distinguish them.

Next, we consider the $\mathcal{N}=4$ superconformal Chern-Simons theories described by circular quiver diagrams which are the three-dimensional worldvolume theories of D3-branes along $012$ directions. D3-branes on each interval give rise to a vector multiplet of $U\left(N_{a}\right)$ group, while each 5-brane gives rise to two bifundamental hypermultiplets. The Chern-Simons level on each interval is
\begin{equation}
k_{a}=\frac{k}{2}\left(s_{a}-s_{a-1}\right).\label{eq:kDef}
\end{equation}
One important observable of these supersymmetric gauge theories is the partition functions. Localization technique \cite{Kapustin:2009kz} reduces the path integral computing the partition functions on round three-sphere to matrix models. These matrix models are parameterized by the Cartan subalgebra $\alpha_{n}^{\left(a\right)}\left(n=1,2,\ldots,N_{a}\right)$ of each $\mathrm{U}\left(N_{a}\right)$ group:
\begin{equation}
Z_{k}^{\bm{s}}\left(\bm{N}\right)=\prod_{a}\left(\frac{1}{N_{a}!}\int d^{N_{a}}\alpha^{\left(a\right)}\right)Z_{\mathrm{classic}}Z_{\mathrm{1-loop}}^{\mathrm{vector}}Z_{\mathrm{1-loop}}^{\mathrm{hyper}}.
\end{equation}
$Z_{\mathrm{classic}}$ captures the value of the action, while $Z_{\mathrm{1-loop}}^{\mathrm{vector}}$ and $Z_{\mathrm{1-loop}}^{\mathrm{hyper}}$ capture the 1-loop determinants of vector multiplets and hypermultiplets respectively. $N_{a}!$ is the order of the Weyl group of $\mathrm{U}\left(N_{a}\right)$.

$Z_{\mathrm{classic}}$ contains the contributions from Chern-Simons terms. Explicitly, the contribution is
\begin{equation}
e^{S_{k_{a}}\left(N_{a};\alpha^{\left(a\right)}\right)},\label{eq:Classical}
\end{equation}
for each $\left\langle \star N_{a}\star\right\rangle $, where
\begin{equation}
S_{k}\left(N;\alpha\right)=i\pi k\sum_{n}^{N}\alpha_{n}^{2}.
\end{equation}
The contribution to $Z_{\mathrm{1-loop}}^{\mathrm{vector}}$ is
\begin{equation}
\left(\prod_{n<n'}^{N_{a}}2\sinh\pi\left(\alpha_{n}^{\left(a\right)}-\alpha_{n'}^{\left(a\right)}\right)\right)^{2},\label{eq:VectorOneLoop}
\end{equation}
for each $\left\langle \star N_{a}\star\right\rangle $, and the contribution to $Z_{\mathrm{1-loop}}^{\mathrm{hyper}}$ is
\begin{equation}
\frac{1}{\prod_{m}^{N_{a}}\prod_{n}^{N_{a+1}}2\cosh\pi\left(\alpha_{m}^{\left(a\right)}-\alpha_{n}^{\left(a+1\right)}\right)},\label{eq:HyperOneLoop}
\end{equation}
for each $\left\langle N_{a}\star N_{a+1}\right\rangle $. Therefore, the contributions of the 1-loop determinants are
\begin{align}
Z\left(N_{a},N_{a+1};\alpha^{\left(a\right)},\alpha^{\left(a+1\right)}\right) & =\frac{\prod_{m<m'}^{N_{a}}2\sinh\pi\left(\alpha_{m}^{\left(a\right)}-\alpha_{m'}^{\left(a\right)}\right)\prod_{n<n'}^{N_{a+1}}2\sinh\pi\left(\alpha_{n}^{\left(a+1\right)}-\alpha_{n'}^{\left(a+1\right)}\right)}{\prod_{m}^{N_{a}}\prod_{n}^{N_{a+1}}2\cosh\pi\left(\alpha_{m}^{\left(a\right)}-\alpha_{n}^{\left(a+1\right)}\right)},\label{eq:OneLoop}
\end{align}
for each $\left\langle N_{a}\star N_{a+1}\right\rangle $. In summary, the partition functions of $\mathcal{N}=4$ Chern-Simons theories labeled by $\bm{s}$ and $\bm{N}$ are
\begin{align}
Z_{k}^{\boldsymbol{s}}\left(\boldsymbol{N}\right) & =\left(\prod_{a}^{r}\frac{i^{-\frac{1}{2}\mathrm{sgn}\left(k_{a}\right)N_{a}^{2}}}{N_{a}!}\right)\int\left(\prod_{a}^{r}\frac{d^{N_{a}}\alpha^{\left(a\right)}}{\hbar^{N_{a}}}e^{S_{k_{a}}\left(N_{a};\frac{\alpha^{\left(a\right)}}{\hbar}\right)}\right)\prod_{a}^{r}Z\left(N_{a},N_{a+1};\frac{\alpha^{\left(a\right)}}{\hbar},\frac{\alpha^{\left(a+1\right)}}{\hbar}\right),\label{eq:PFall}
\end{align}
where $N_{r+1}=N_{1}$, $\alpha^{\left(r+1\right)}=\alpha^{\left(1\right)}$, $\hbar=2\pi k$ and
\begin{equation}
\mathrm{sgn}\left(k\right)=\begin{cases}
1 & \left(k>0\right)\\
0 & \left(k=0\right)\\
-1 & \left(k<0\right)
\end{cases}.
\end{equation}
For later convenience, we changed the integration variables as $\alpha^{\left(a\right)}\rightarrow\frac{\alpha^{\left(a\right)}}{\hbar}$. The definition of phase which we adopt here can be considered to be a natural extension of the ABJ matrix model \cite{Drukker:2010nc}. This definition is also natural from the point of view of our results.

\subsection{Strategy of computation\label{subsec:Strategy}}

What we will achieve in this paper is to rewrite the matrix models \eqref{eq:PFall} as the partition functions of ideal Fermi gases with $N$ particles:
\begin{equation}
Z_{k}^{\boldsymbol{s}}\left(\boldsymbol{N}\right)=\frac{C_{k,\boldsymbol{M}}^{\boldsymbol{s}}}{N!}\int d^{N}\alpha\det\left(\left[\bra{\alpha_{\overline{n}}}\widehat{\rho}_{\boldsymbol{M}}^{\boldsymbol{s}}\left(\widehat{q},\widehat{p}\right)\ket{\alpha_{n}}\right]_{\overline{n},n}^{N\times N}\right).\label{eq:FGFform}
\end{equation}
This idea is called Fermi gas formalism \cite{Marino:2011eh}. The factor $C_{k,\boldsymbol{M}}^{\boldsymbol{s}}$ does not depend on $N$. $\widehat{q}$ and $\widehat{p}$ are Hermitian operators and are a position operator and a momentum operator respectively. They satisfy canonical commutation relation $\left[\widehat{q},\widehat{p}\right]=i\hbar=2\pi ik$. $\ket{\cdot}$ denotes a position eigenvector. We also introduce a symbol $\kket{\cdot}$ denoting a momentum eigenvector. The inner products of these vectors are
\begin{align}
 & \braket{q_{1}|q_{2}}=\delta\left(q_{1}-q_{2}\right),\quad\bbrakket{p_{1}|p_{2}}=\delta\left(p_{1}-p_{2}\right),\nonumber \\
 & \brakket{q|p}=\frac{\sqrt{k}}{\hbar}e^{\frac{i}{\hbar}pq},\quad\bbraket{p|q}=\frac{\sqrt{k}}{\hbar}e^{-\frac{i}{\hbar}pq}.\label{eq:Normalization}
\end{align}
$\widehat{\rho}_{\boldsymbol{M}}^{\boldsymbol{s}}$ is the one-particle density matrix of the Fermions. The density matrix depends on $k$ although we omit its subscript. Note that the density matrix can be determined only up to similarity transformations because the values of the matrix models are invariant under the similarity transformations as you can easily see from \eqref{eq:FGFform}. In the rest of this section, we explain our strategy to apply the Fermi gas formalism to the matrix models.

The important idea for this purpose is to cut the matrix model \eqref{eq:PFall} at positions where corresponding ranks of nodes are lowest and apply the Fermi gas formalism to each part locally. In other words, we focus on each lump of the 5-brane factors \eqref{eq:OneLoop} (and the Fresnel factors) with ranks deformed from the lowest rank, hence we call these parts deformed factors. In terms of the brane configurations, this procedure of cutting means that we cut D3-branes at positions where the number of D3-branes is lowest. In order to achieve this strategy, we first explicitly define the deformed factors. We start with the separated parts of brane configurations:
\begin{equation}
\left\langle 0\star M_{1}\star M_{2}\star\cdots\star M_{r}\star0\right\rangle _{N}.\label{eq:PartConfig}
\end{equation}
We also parameterize the sequence of 5-branes as $\boldsymbol{s}=\left(s_{0},s_{1},\ldots,s_{r}\right)$. The corresponding parts of matrix models, which we call deformed factors, are obtained by applying the translation rule \eqref{eq:Classical} and \eqref{eq:OneLoop}:
\begin{align}
 & Z_{k,\boldsymbol{M}}^{\boldsymbol{s}}\left(N;\frac{\alpha}{\hbar},\frac{\beta}{\hbar}\right)\nonumber \\
 & =\left(\prod_{a}^{r}\frac{i^{-\frac{1}{2}\mathrm{sgn}\left(k_{a}\right)\left\{ \left(N+M_{a}\right)^{2}-N^{2}\right\} }}{\left(N+M_{a}\right)!}\right)\nonumber \\
 & \,\,\,\times\int\left(\prod_{a}^{r}\frac{d^{N+M_{a}}\alpha^{\left(a\right)}}{\hbar^{N+M_{a}}}e^{S_{k_{a}}\left(N+M_{a};\frac{\alpha^{\left(a\right)}}{\hbar}\right)}\right)\prod_{a=0}^{r}Z\left(N+M_{a},N+M_{a+1};\frac{\alpha^{\left(a\right)}}{\hbar},\frac{\alpha^{\left(a+1\right)}}{\hbar}\right),\label{eq:PartGenDef}
\end{align}
where $\alpha^{\left(0\right)}=\alpha$, $\alpha^{\left(r+1\right)}=\beta$ and $M_{0}=M_{r+1}=0$. $\alpha$ and $\beta$ are variables respectively corresponding to the leftmost and the rightmost D3-branes in the brane configuration \eqref{eq:PartConfig}. We multiplied the deformed factors by the phase factor $i^{\frac{1}{2}\mathrm{sgn}\left(k_{a}\right)N^{2}}$. This factor is canceled in the whole matrix models because of the definition of $k_{a}$ in \eqref{eq:kDef}. Note that when $N=0$, the deformed factors are equal to the matrix models computing the partition function of the worldvolume theories of brane configurations \eqref{eq:PartConfig} with $N=0$. These worldvolume theories are $\mathcal{N}=4$ superconformal Chern-Simons theories described by linear quiver diagrams \cite{Gaiotto:2008sd,Hosomichi:2008jd,Imamura:2008dt}.

The whole matrix models \eqref{eq:PFall} are obtained by gluing the deformed factors with integration while keeping in mind the Fresnel factors \eqref{eq:Classical}. Concretely, let us cut brane configurations as
\begin{equation}
\boldsymbol{s}=\left(\boldsymbol{s}^{1},\boldsymbol{s}^{2},\ldots,\boldsymbol{s}^{R}\right).
\end{equation}
As explained above, all the ranks between $\boldsymbol{s}^{A-1}$ and $\boldsymbol{s}^{A}$ are the lowest rank $N$. The whole matrix models are
\begin{equation}
Z_{k}^{\boldsymbol{s}}\left(\boldsymbol{N}\right)=\frac{1}{\left(N!\right)^{R}}\int\left(\prod_{A}^{R}\frac{d^{N}\alpha^{\left(A\right)}}{\hbar^{N}}e^{S_{k_{A}}\left(N;\frac{\alpha^{\left(A\right)}}{\hbar}\right)}\right)\prod_{A}^{R}Z_{k,\boldsymbol{M}^{A}}^{\boldsymbol{s}^{A}}\left(N;\frac{\alpha^{\left(A\right)}}{\hbar},\frac{\alpha^{\left(A+1\right)}}{\hbar}\right),\label{eq:PFdecom}
\end{equation}
where $\alpha^{\left(R+1\right)}=\alpha^{\left(1\right)}$, $k_{A}=\frac{k}{2}\left(s_{0}^{A}-s_{r_{A-1}}^{A-1}\right)$ ($s_{r_{A-1}}^{A-1}$ is the last element of $\boldsymbol{s}^{A-1}$), and $\boldsymbol{M}^{A}$ denotes the relative ranks in $\boldsymbol{s}^{A}$.

At this point, we have divided the matrix models into the deformed factors. Next, we explain how we rewrite the matrix models as the partition functions of the ideal Fermi gases \eqref{eq:FGFform}. First, we incorporate all the Fresnel factors into deformed factors $Z_{k,\boldsymbol{M}^{A}}^{\boldsymbol{s}^{A}}$ and transform them consistently between the nearest neighbors, and we denote by $\widetilde{Z}_{k,\boldsymbol{M}^{A}}^{\boldsymbol{s}^{A}}$ the results (we also call them deformed factors). We explain the detail of this procedure and definition of $\widetilde{Z}_{k,\boldsymbol{M}^{A}}^{\boldsymbol{s}^{A}}$ soon later. After this operation, the whole matrix models become
\begin{equation}
Z_{k}^{\boldsymbol{s}}\left(\boldsymbol{N}\right)=\frac{1}{\left(N!\right)^{R}}\int\left(\prod_{A}^{R}\frac{d^{N}\alpha^{\left(A\right)}}{\hbar^{N}}\right)\prod_{A}^{R}\widetilde{Z}_{k,\boldsymbol{M}^{A}}^{\boldsymbol{s}^{A}}\left(N;\frac{\alpha^{\left(A\right)}}{\hbar},\frac{\alpha^{\left(A+1\right)}}{\hbar}\right).\label{eq:PFsim}
\end{equation}
If we can now put all $\widetilde{Z}_{k,\boldsymbol{M}^{A}}^{\boldsymbol{s}^{A}}$ into the following form,
\begin{equation}
\widetilde{Z}_{k,\boldsymbol{M}^{A}}^{\boldsymbol{s}^{A}}\left(N;\frac{\alpha^{\left(A\right)}}{\hbar},\frac{\alpha^{\left(A+1\right)}}{\hbar}\right)=C_{k,\boldsymbol{M}^{A}}^{\boldsymbol{s}^{A}}\hbar^{N}\det\left(\left[\bra{\alpha_{\overline{n}}^{\left(A\right)}}\widehat{\rho}_{\boldsymbol{M}^{A}}^{\boldsymbol{s}^{A}}\ket{\alpha_{n}^{\left(A+1\right)}}\right]_{\overline{n},n}^{N\times N}\right),\label{eq:ZtildeGenRes}
\end{equation}
then using the identity
\begin{equation}
\frac{1}{N!}\int d^{N}\beta\det\left(\left[\bra{\alpha_{\overline{n}}}\widehat{\rho}_{1}\ket{\beta_{n}}\right]_{\overline{n},n}^{N\times N}\right)\det\left(\left[\bra{\beta_{\overline{n}}}\widehat{\rho}_{2}\ket{\gamma_{n}}\right]_{\overline{n},n}^{N\times N}\right)=\det\left(\left[\bra{\alpha_{\overline{n}}}\widehat{\rho}_{1}\widehat{\rho}_{2}\ket{\gamma_{n}}\right]_{\overline{n},n}^{N\times N}\right),\label{eq:DetGlue}
\end{equation}
we finally obtain
\begin{equation}
Z_{k}^{\bm{s}}\left(\bm{N}\right)=\frac{\prod_{A}^{R}C_{k,\boldsymbol{M}^{A}}^{\boldsymbol{s}^{A}}}{N!}\int d^{N}\alpha\det\left(\left[\bra{\alpha_{\overline{n}}}\prod_{A}^{R}\widehat{\rho}_{\boldsymbol{M}^{A}}^{\boldsymbol{s}^{A}}\ket{\alpha_{n}}\right]_{\overline{n},n}^{N\times N}\right).\label{eq:PFallRes}
\end{equation}
This is clearly the form of \eqref{eq:FGFform}, and therefore we have completed applying the Fermi gas formalism to the matrix models.

The above strategy has the following advantage. If we can put $\widetilde{Z}_{k,\boldsymbol{M}^{A}}^{\boldsymbol{s}^{A}}$ into \eqref{eq:ZtildeGenRes}, then by gluing those results we obtain more complex matrix models, namely the matrix models corresponding to a larger number of 5-branes (although the number of glued D3-branes should be the lowest). Therefore, we can apply the Fermi gas formalism to a wide class of rank deformed matrix models by using our separate computation approach.

We now complete the explanation of our strategy of the computation by explaining how we transform $Z_{k,\boldsymbol{M}^{A}}^{\boldsymbol{s}^{A}}$ into $\widetilde{Z}_{k,\boldsymbol{M}^{A}}^{\boldsymbol{s}^{A}}$. Let us return to the whole matrix models \eqref{eq:PFall}. The determinant formula \eqref{eq:CauchyDet} plays an important role because this formula puts $Z\left(N_{a},N_{a+1};\frac{\alpha}{\hbar},\frac{\beta}{\hbar}\right)$ into the form that all elements are the inner product of $\bra{\alpha}$ and $\ket{\beta}$ or constant vectors. Because we are considering circular (or linear) quiver diagrams, all the integrals are now the following form:
\begin{equation}
\int d\alpha e^{S_{k_{a}}\left(N_{a};\frac{\alpha^{\left(a\right)}}{\hbar}\right)}\ket{\alpha^{\left(\alpha\right)}}\bra{\alpha^{\left(\alpha\right)}}=\int d\alpha e^{\frac{i\mathrm{sgn}\left(k_{a}\right)}{2\hbar}\left(\alpha^{\left(\alpha\right)}\right)^{2}}\ket{\alpha^{\left(\alpha\right)}}\bra{\alpha^{\left(\alpha\right)}}.
\end{equation}
Let us focus on the definition of the Chern-Simons levels $k_{a}=\frac{k}{2}\left(s_{a}-s_{a-1}\right)$. We realize that we can transform the position eigenvectors into the following form depending on the values of $\left(s_{a-1},s_{a}\right)$:
\begin{align}
\int d\alpha\ket{\alpha}\bra{\alpha}=\int d\alpha e^{\frac{i}{2\hbar}\widehat{p}^{2}}\ket{\alpha}\bra{\alpha}e^{-\frac{i}{2\hbar}\widehat{p}^{2}} & \quad:\left(+1,+1\right),\nonumber \\
\int d\alpha e^{\frac{i}{2\hbar}\alpha^{2}}\ket{\alpha}\bra{\alpha}=\int d\alpha e^{\frac{i}{2\hbar}\widehat{q}^{2}}e^{\frac{i}{2\hbar}\widehat{p}^{2}}\ket{\alpha}\bra{\alpha}e^{-\frac{i}{2\hbar}\widehat{p}^{2}} & \quad:\left(-1,+1\right),\nonumber \\
\int d\alpha e^{-\frac{i}{2\hbar}\alpha^{2}}\ket{\alpha}\bra{\alpha}=\int d\alpha e^{\frac{i}{2\hbar}\widehat{p}^{2}}\ket{\alpha}\bra{\alpha}e^{-\frac{i}{2\hbar}\widehat{p}^{2}}e^{-\frac{i}{2\hbar}\widehat{q}^{2}} & \quad:\left(+1,-1\right),\nonumber \\
\int d\alpha\ket{\alpha}\bra{\alpha}=\int d\alpha e^{\frac{i}{2\hbar}\widehat{q}^{2}}e^{\frac{i}{2\hbar}\widehat{p}^{2}}\ket{\alpha}\bra{\alpha}e^{-\frac{i}{2\hbar}\widehat{p}^{2}}e^{-\frac{i}{2\hbar}\widehat{q}^{2}} & \quad:\left(-1,-1\right).
\end{align}
The key point is that each position eigenvector in $s=\pm1$ sector has the same transformation rule:
\begin{align}
\bra{\alpha}\rightarrow\bra{\alpha}e^{-\frac{i}{2\hbar}\widehat{p}^{2}},\quad\ket{\alpha}\rightarrow e^{\frac{i}{2\hbar}\widehat{p}^{2}}\ket{\alpha} & \quad:s=+1,\nonumber \\
\bra{\alpha}\rightarrow\bra{\alpha}e^{-\frac{i}{2\hbar}\widehat{p}^{2}}e^{-\frac{i}{2\hbar}\widehat{q}^{2}},\quad\ket{\alpha}\rightarrow e^{\frac{i}{2\hbar}\widehat{q}^{2}}e^{\frac{i}{2\hbar}\widehat{p}^{2}}\ket{\alpha} & \quad:s=-1.\label{eq:BraketChange}
\end{align}
Therefore, the simultaneous operations of eliminating all the Fresnel factors and transforming all $Z\left(N_{a},N_{a+1};\frac{\alpha}{\hbar},\frac{\beta}{\hbar}\right)$ according to \eqref{eq:BraketChange} give the correct transformation of the whole matrix models \eqref{eq:PFall}. This is the same for our separate computation approach. Hence, whenever we compute a deformed factor $Z_{k,\boldsymbol{M}}^{\boldsymbol{s}}$ locally, we transform the deformed factor according to \eqref{eq:BraketChange}, and we denote by $\widetilde{Z}_{k,\boldsymbol{M}}^{\boldsymbol{s}}$ the result. After these operations, the whole matrix models have been the form \eqref{eq:PFsim}, so that we can accomplish our computation as explained if we can put $\widetilde{Z}_{k,\boldsymbol{M}}^{\boldsymbol{s}}$ into the form \eqref{eq:ZtildeGenRes}. Note that the technical advantages of \eqref{eq:BraketChange} is that each determinant captures simple similarity transformation generated by $e^{\frac{i}{2\hbar}\widehat{p}^{2}}$ and $e^{\frac{i}{2\hbar}\widehat{q}^{2}}e^{\frac{i}{2\hbar}\widehat{p}^{2}}$ for $s=+1$ and $s=-1$, respectively.

In summary, what we will achieve in this paper is as follows. We study a special class of deformed factors \eqref{eq:PartGenDef} which we will explain in section \ref{subsec:Target}. We first use the determinant formula \eqref{eq:CauchyDet}, so that all elements become the inner product of $\bra{\alpha}$ and $\ket{\beta}$ or constant vectors. We then eliminate all the Fresnel factors and transform the position eigenvectors according to \eqref{eq:BraketChange}, and the resulting deformed factors are denoted by $\widetilde{Z}_{k,\boldsymbol{M}}^{\boldsymbol{s}}$. We finally transform them into the form \eqref{eq:ZtildeGenRes}.

\subsection{Simple examples}

In this section we review the computation for deformed factors associated with the brane configurations consisting of less than three 5-branes, and we confirm our strategy explained in the previous section. We first review one 5-brane case in section \ref{subsec:One-5-brane}. The deformed factor consisting of one NS5-brane and one $\left(1,k\right)$5-brane is closely related to the ABJ theory, hence we review this deformed factor from this viewpoint in section \ref{subsec:ABJcase}.

\subsubsection{One 5-brane\label{subsec:One-5-brane}}

The computation of the matrix models associated with the $\mathcal{N}=4$ Chern-Simons theories described by circular quiver diagrams naturally reduces to the local computation for each 5-brane if all the nodes have equal ranks \cite{Marino:2011eh}. Therefore, in this case, it is enough to compute the deformed factors of $\left(s_{a}\right)=\pm1$. The corresponding brane configurations \eqref{eq:PartConfig} are
\begin{equation}
\left\langle 0\bullet0\right\rangle _{N},\quad\left\langle 0\circ0\right\rangle _{N},\label{eq:ConfiguTriv}
\end{equation}
respectively, and the corresponding deformed factors \eqref{eq:PartGenDef} are
\begin{equation}
Z_{k}^{\left(\pm1\right)}\left(N;\frac{\alpha}{\hbar},\frac{\beta}{\hbar}\right)=\frac{\prod_{\overline{n}<\overline{n}'}^{N}2\sinh\frac{\alpha_{\overline{n}}-\alpha_{\overline{n}'}}{2k}\prod_{n<n'}^{N}2\sinh\frac{\beta_{n}-\beta_{n'}}{2k}}{\prod_{\overline{n}}^{N}\prod_{n}^{N}2\cosh\frac{\alpha_{\overline{n}}-\beta_{n}}{2k}}.
\end{equation}
This expression can be written in a different way using the determinant formula \eqref{eq:CauchyDet} as
\begin{equation}
Z_{k}^{\left(\pm1\right)}\left(N;\frac{\alpha}{\hbar},\frac{\beta}{\hbar}\right)=\hbar^{N}\det\left(\left[\bra{\alpha_{\overline{n}}}\frac{1}{2\cosh\frac{\widehat{p}}{2}}\ket{\beta_{n}}\right]_{\overline{n},n}^{N\times N}\right).
\end{equation}
We transform these deformed factors into
\begin{align}
\widetilde{Z}^{\left(+1\right)}\left(N;\frac{\alpha}{\hbar},\frac{\beta}{\hbar}\right) & =\hbar^{N}\det\left(\left[\bra{\alpha_{\overline{n}}}e^{-\frac{i}{2\hbar}\widehat{p}^{2}}\frac{1}{2\cosh\frac{\widehat{p}}{2}}e^{\frac{i}{2\hbar}\widehat{p}^{2}}\ket{\beta_{n}}\right]_{\overline{n},n}^{N\times N}\right),\nonumber \\
\widetilde{Z}^{\left(-1\right)}\left(N;\frac{\alpha}{\hbar},\frac{\beta}{\hbar}\right) & =\hbar^{N}\det\left(\left[\bra{\alpha_{\overline{n}}}e^{-\frac{i}{2\hbar}\widehat{p}^{2}}e^{-\frac{i}{2\hbar}\widehat{q}^{2}}\frac{1}{2\cosh\frac{\widehat{p}}{2}}e^{\frac{i}{2\hbar}\widehat{q}^{2}}e^{\frac{i}{2\hbar}\widehat{p}^{2}}\ket{\beta_{n}}\right]_{\overline{n},n}^{N\times N}\right),
\end{align}
according to \eqref{eq:BraketChange}. This similarity transformation do not affect for $s=+1$ case. On the other hand, for $s=-1$ case, this similarity transformation drastically changes the components. Using the identity
\begin{equation}
e^{-\frac{i}{2\hbar}\widehat{p}^{2}}e^{-\frac{i}{2\hbar}\widehat{q}^{2}}f\left(\widehat{p}\right)e^{\frac{i}{2\hbar}\widehat{q}^{2}}e^{\frac{i}{2\hbar}\widehat{p}^{2}}=f\left(\widehat{q}\right),\label{eq:Simptoq}
\end{equation}
we finally obtain
\begin{align}
\widetilde{Z}^{\left(+1\right)}\left(N;\frac{\alpha}{\hbar},\frac{\beta}{\hbar}\right) & =\hbar^{N}\det\left(\left[\bra{\alpha_{\overline{n}}}\frac{1}{2\cosh\frac{\widehat{p}}{2}}\ket{\beta_{n}}\right]_{\overline{n},n}^{N\times N}\right),\nonumber \\
\widetilde{Z}^{\left(-1\right)}\left(N;\frac{\alpha}{\hbar},\frac{\beta}{\hbar}\right) & =\hbar^{N}\det\left(\left[\bra{\alpha_{\overline{n}}}\frac{1}{2\cosh\frac{\widehat{q}}{2}}\ket{\beta_{n}}\right]_{\overline{n},n}^{N\times N}\right).\label{eq:ZtrivRes}
\end{align}
This form is clearly \eqref{eq:ZtildeGenRes}. Therefore, we have succeeded in applying the Fermi gas formalism. Note that the above expression shows that
\begin{equation}
\widehat{\rho}^{\left(+1\right)}=\frac{1}{2\cosh\frac{\widehat{p}}{2}},\quad\widehat{\rho}^{\left(-1\right)}=\frac{1}{2\cosh\frac{\widehat{q}}{2}}.\label{eq:rhoTrivRes}
\end{equation}

\subsubsection{ABJ theory\label{subsec:ABJcase}}

In \cite{Awata:2012jb,Honda:2013pea,Honda:2014npa}, they applied the Fermi gas formalism to the ABJ matrix model with complicated computations. After that, it was gradually realized that this type of computations can be performed more systematically by using operator formalism \cite{Moriyama:2015asx,Moriyama:2016xin,Moriyama:2016kqi,Kiyoshige:2016lno,Moriyama:2017gye}. In \cite{Kubo:2019ejc}, by using this formalism, they computed the deformed factor consisting of one NS5-brane and one $\left(1,k\right)$5-brane, which is closely related to the ABJ theory. In this section we review this computation for the ease of understanding the main computation of this paper in section \ref{subsec:Calculation} where we will apply the Fermi gas formalism to more complicated deformed factors.

The brane configuration of the ABJ theory is $\left\langle 0\bullet M\circ\right\rangle _{N}^{\mathrm{P}}$. We restrict the number of D3-branes ending on an NS5-brane and a $\left(1,k\right)$5-brane to $M\geq0$ and equal to or less than $k$,\footnote{It was argued in \cite{Aharony:2008gk} that the ABJ theory with $M>k$ do not exist as an unitary theory.} namely
\begin{equation}
M\leq k.\label{eq:ConfigCondABJ}
\end{equation}
The part of the brane configuration needed for the ABJ theory is $\left\langle 0\bullet M\circ0\right\rangle _{N}$, and corresponding deformed factor \eqref{eq:PartGenDef} is
\begin{align}
 & Z_{k,M}^{\left(+1,-1\right)}\left(N;\frac{\alpha}{\hbar},\frac{\beta}{\hbar}\right)\nonumber \\
 & =\frac{i^{\frac{1}{2}\left\{ \left(N+M\right)^{2}-N^{2}\right\} }}{\left(N+M\right)!}\int\frac{d^{N+M}\nu}{\hbar^{N+M}}e^{-\frac{i}{2\hbar}\sum_{m}^{N+M}\nu_{m}^{2}}\nonumber \\
 & \,\,\,\times\frac{\prod_{n<n'}^{N}2\sinh\frac{\alpha_{n}-\alpha_{n'}}{2k}\prod_{m<m'}^{N+M}2\sinh\frac{\nu_{m}-\nu_{m'}}{2k}}{\prod_{n}^{N}\prod_{m}^{N+M}2\cosh\frac{\alpha_{n}-\nu_{m}}{2k}}\frac{\prod_{m<m'}^{N+M}2\sinh\frac{\nu_{m}-\nu_{m'}}{2k}\prod_{n<n'}^{N}2\sinh\frac{\beta_{n}-\beta_{n'}}{2k}}{\prod_{m}^{N+M}\prod_{n}^{N}2\cosh\frac{\nu_{m}-\beta_{n}}{2k}}.\label{eq:ZpmDef}
\end{align}
Using this deformed factor, the ABJ matrix model can be expressed as
\begin{equation}
Z_{k}^{\left(+1,-1\right)}\left(N,N+M\right)=\frac{1}{N!}\int\frac{d^{N}\mu}{\hbar^{N}}e^{\frac{i}{2\hbar}\sum_{n}^{N}\mu_{n}^{2}}Z_{k,M}^{\left(+1,-1\right)}\left(N;\frac{\mu}{\hbar},\frac{\mu}{\hbar}\right).\label{eq:ZABJdef}
\end{equation}

We start computing the deformed factor \eqref{eq:ZpmDef}. Using the determinant formula \eqref{eq:CauchyDet} we get
\begin{align}
Z_{k,M}^{\left(+1,-1\right)}\left(N;\frac{\alpha}{\hbar},\frac{\beta}{\hbar}\right) & =\frac{i^{\frac{1}{2}\left\{ \left(N+M\right)^{2}-N^{2}\right\} }}{\left(N+M\right)!}\int\frac{d^{N+M}\nu}{\hbar^{N+M}}e^{-\frac{i}{2\hbar}\sum_{m}^{N+M}\nu_{m}^{2}}\nonumber \\
 & \,\,\,\times\det\left(\begin{array}{c}
\left[\hbar\bra{\alpha_{n}}\frac{1}{2\cosh\frac{\widehat{p}-i\pi M}{2}}\ket{\nu_{m}}\right]_{n,m}^{N\times\left(N+M\right)}\\
\left[\frac{\hbar}{\sqrt{k}}\bbraket{2\pi it_{M,j}|\nu_{m}}\right]_{j,m}^{M\times\left(N+M\right)}
\end{array}\right)\nonumber \\
 & \,\,\,\times\det\left(\begin{array}{cc}
\left[\hbar\bra{\nu_{m}}\frac{1}{2\cosh\frac{\widehat{p}+i\pi M}{2}}\ket{\beta_{n}}\right]_{m,n}^{_{\times N}^{\left(N+M\right)}} & \left[\frac{\hbar}{\sqrt{k}}\brakket{\nu_{m}|-2\pi it_{M,j}}\right]_{m,j}^{_{\times M}^{\left(N+M\right)}}\end{array}\right),\label{SimEx1}
\end{align}
where
\begin{equation}
t_{M,j}=\frac{M+1}{2}-j.\label{eq:tMdef}
\end{equation}
We transform all position eigenvectors according to \eqref{eq:BraketChange}, namely,
\begin{align}
\bra{\alpha_{n}} & \rightarrow\bra{\alpha_{n}}e^{-\frac{i}{2\hbar}\widehat{p}^{2}},\quad\ket{\nu_{m}}\rightarrow e^{\frac{i}{2\hbar}\widehat{p}^{2}}\ket{\nu_{m}},\nonumber \\
\bra{\nu_{m}} & \rightarrow\bra{\nu_{m}}e^{-\frac{i}{2\hbar}\widehat{p}^{2}}e^{-\frac{i}{2\hbar}\widehat{q}^{2}},\quad\ket{\beta_{n}}\rightarrow e^{\frac{i}{2\hbar}\widehat{q}^{2}}e^{\frac{i}{2\hbar}\widehat{p}^{2}}\ket{\beta_{n}},
\end{align}
and we eliminate the Fresnel factors $e^{-\frac{i}{2\hbar}\sum_{m}^{N+M}\nu_{m}^{2}}$. The result is denoted by $\widetilde{Z}_{k,M}^{\left(+1,-1\right)}$. As explained in section \ref{subsec:Strategy}, what we operated for $\nu_{m}$ is just
\begin{equation}
\int d\nu_{m}e^{-\frac{i}{2\hbar}\nu_{m}^{2}}\ket{\nu_{m}}\bra{\nu_{m}}=\int d\nu_{m}e^{\frac{i}{2\hbar}\widehat{p}^{2}}\ket{\nu_{m}}\bra{\nu_{m}}e^{-\frac{i}{2\hbar}\widehat{p}^{2}}e^{-\frac{i}{2\hbar}\widehat{q}^{2}},
\end{equation}
and the ABJ matrix model \eqref{eq:ZABJdef} is expressed as
\begin{equation}
Z_{k}^{\left(+1,-1\right)}\left(N,N+M\right)=\frac{1}{N!}\int\frac{d^{N}\mu}{\hbar^{N}}\widetilde{Z}_{k,M}^{\left(+1,-1\right)}\left(N;\frac{\mu}{\hbar},\frac{\mu}{\hbar}\right).\label{eq:ZABJ1}
\end{equation}
We apply \eqref{eq:Simptoq} and the following identity:\footnote{The second identity is for later convenience.}
\begin{equation}
e^{-\frac{i}{2\hbar}\widehat{p}^{2}}e^{-\frac{i}{2\hbar}\widehat{q}^{2}}\kket p=\frac{1}{\sqrt{i}}e^{\frac{i}{2\hbar}p^{2}}\ket p,\quad\bbra pe^{\frac{i}{2\hbar}\widehat{q}^{2}}e^{\frac{i}{2\hbar}\widehat{p}^{2}}=\sqrt{i}e^{-\frac{i}{2\hbar}p^{2}}\bra p,\label{eq:Vecptoq}
\end{equation}
to $\widetilde{Z}_{k,M}^{\left(+1,-1\right)}$, so that we obtain
\begin{align}
\widetilde{Z}_{k,M}^{\left(+1,-1\right)}\left(N;\frac{\alpha}{\hbar},\frac{\beta}{\hbar}\right) & =\frac{i^{\frac{1}{2}\left\{ \left(N+M\right)^{2}-N^{2}\right\} -\frac{M}{2}}e^{i\theta_{k,M}}}{\left(N+M\right)!}k^{-\frac{M}{2}}\int d^{N+M}\nu\nonumber \\
 & \,\,\,\times\det\left(\begin{array}{c}
\left[\hbar\bra{\alpha_{n}}\frac{1}{2\cosh\frac{\widehat{p}-i\pi M}{2}}\ket{\nu_{m}}\right]_{n,m}^{N\times\left(N+M\right)}\\
\left[\frac{\hbar}{\sqrt{k}}\bbraket{2\pi it_{M,j}|\nu_{m}}\right]_{j,m}^{M\times\left(N+M\right)}
\end{array}\right)\nonumber \\
 & \,\,\,\times\det\left(\begin{array}{cc}
\left[\bra{\nu_{m}}\frac{1}{2\cosh\frac{\widehat{q}+i\pi M}{2}}\ket{\beta_{n}}\right]_{m,n}^{_{\times N}^{\left(N+M\right)}} & \left[\braket{\nu_{m}|-2\pi it_{M,j}}\right]_{m,j}^{_{\times M}^{\left(N+M\right)}}\end{array}\right),
\end{align}
where
\begin{equation}
\theta_{k,M}=-\frac{\pi}{6k}\left(M^{3}-M\right).\label{eq:phaseABJ}
\end{equation}
This phase factor comes from both constant vectors with the help of
\begin{equation}
\frac{i}{2\hbar}\prod_{j}^{M}\left(\pm2\pi it_{M,j}\right)^{2}=\frac{i}{2}\theta_{k,M}.
\end{equation}
We use the following identity:\footnote{For later convenience, we write general form. In this case, $f_{mj}'=0$.}
\begin{align}
 & \frac{1}{N!}\int d^{N}\alpha\det\left(\begin{array}{cc}
\left[f_{m}\left(\alpha_{n}\right)\right]_{m,n}^{\left(N+M\right)\times N} & \left[f_{mj}'\right]_{m,j}^{\left(N+M\right)\times M}\end{array}\right)\det\left(\left[g_{n'}\left(\alpha_{n}\right)\right]_{n,n'}^{N\times N}\right)\nonumber \\
 & =\int d^{N}\alpha\det\left(\begin{array}{cc}
\left[f_{m}\left(\alpha_{n}\right)\right]_{m,n}^{\left(N+M\right)\times N} & \left[f_{mj}'\right]_{m,j}^{\left(N+M\right)\times M}\end{array}\right)\prod_{n}^{N}g_{n}\left(\alpha_{n}\right),\label{eq:Diagonalize}
\end{align}
to diagonalize the second determinant, and then we again use the determinant formula \eqref{eq:CauchyDet}. The result is
\begin{align}
\widetilde{Z}_{k,M}^{\left(+1,-1\right)}\left(N;\frac{\alpha}{\hbar},\frac{\beta}{\hbar}\right) & =i^{\frac{1}{2}\left\{ \left(N+M\right)^{2}-N^{2}\right\} -\frac{M}{2}}e^{i\theta_{k,M}}k^{-\frac{M}{2}}\nonumber \\
 & \,\,\,\times\int d^{N+M}\nu\frac{\prod_{n<n'}^{N}2\sinh\frac{\alpha_{n}-\alpha_{n'}}{2k}\prod_{m<m'}^{N+M}2\sinh\frac{\nu_{m}-\nu_{m'}}{2k}}{\prod_{n}^{N}\prod_{m}^{N+M}2\cosh\frac{\alpha_{n}-\nu_{m}}{2k}}\nonumber \\
 & \,\,\,\times\left(\prod_{n}^{N}\bra{\nu_{n}}\frac{1}{2\cosh\frac{\widehat{q}+i\pi M}{2}}\ket{\beta_{n}}\right)\left(\prod_{j}^{M}\braket{\nu_{N+j}|-2\pi it_{M,j}}\right).\label{eq:ZtildeABJ1}
\end{align}
Fortunately, there are direct inner products of position operators, namely the delta functions, in the last line. Actually, it is straightforward to perform the integration over $\nu_{n}$ $\left(1\leq n\leq N\right)$. On the other hand, we should be careful to carry out the integration over $\nu_{N+j}$ $\left(1\leq j\leq M\right)$ because the arguments of the delta functions are now complex number $-2\pi it_{M,j}$. We have to shift the integration contour from $\mathbb{R}$ to $\mathbb{R}-2\pi it_{M,j}$ so that we can use the property of the delta functions. If there are poles in the region where integration contour passes, we have to take account of its residues. Appendix \ref{sec:integral} provides more details.\footnote{\label{fn:Divergence}Strictly speaking, we cannot use the argument in appendix \ref{sec:integral} because the integrand of \eqref{eq:ZtildeABJ1} do not converge at $\nu_{N+j}\rightarrow\pm\infty$. This is because the deformed factors \eqref{eq:PartGenDef} are not convergent. Therefore, in this paper we shall give all Chern-Simons levels a small imaginary part (the sign is chosen such that the integral converges) and taking it to zero in the end.} Nevertheless, we can also perform the remaining integration as usual because the restriction \eqref{eq:ConfigCondABJ} ensures that the integration contour passes through no poles. We confirm this fact. The part of the function of $\nu_{N+j}$ having poles and the region where this part has no poles are
\begin{equation}
\frac{1}{\prod_{n}^{N}2\cosh\frac{\alpha_{n}-\nu_{N+j}}{2k}},\quad\text{No pole region: }\left|\mathrm{Im}\left(\nu_{N+j}\right)\right|<\pi k.
\end{equation}
Moreover, the restriction \eqref{eq:ConfigCondABJ} leads $\left|2\pi t_{M,j}\right|\leq\pi\left(k-1\right)$, so that integration contour passes through no poles as mentioned.

The above argument enable us to perform the integration over $\nu_{m}$, and the result is
\begin{align}
\widetilde{Z}_{k,M}^{\left(+1,-1\right)}\left(N;\frac{\alpha}{\hbar},\frac{\beta}{\hbar}\right) & =i^{\frac{1}{2}\left\{ \left(N+M\right)^{2}-N^{2}\right\} -\frac{M}{2}}e^{i\theta_{k,M}}k^{-\frac{M}{2}}\nonumber \\
 & \,\,\,\times\frac{\prod_{n<n'}^{N}2\sinh\frac{\alpha_{n}-\alpha_{n'}}{2k}\prod_{m<m'}^{N+M}2\sinh\frac{\nu_{m}-\nu_{m'}}{2k}}{\prod_{n}^{N}\prod_{m}^{N+M}2\cosh\frac{\alpha_{n}-\nu_{m}}{2k}}\prod_{n}^{N}\frac{1}{2\cosh\frac{\nu_{n}+i\pi M}{2}},
\end{align}
where
\begin{equation}
\nu_{m}=\begin{cases}
\beta_{m} & \left(1\leq m\leq N\right)\\
-2\pi it_{M,m-N} & \left(N+1\leq m\leq N+M\right)
\end{cases}.\label{eq:RepABJ}
\end{equation}
We divide this into a phase factor depending on $k$, a phase factor $\Omega$ independent of $k$, $Z_{0}$ which is related to the $N$-independent factor $C_{k,\boldsymbol{M}}^{\boldsymbol{s}}$ and $\mathcal{O}$ which is related to the density matrix:
\begin{equation}
\widetilde{Z}_{k,M}^{\left(+1,-1\right)}\left(N;\frac{\alpha}{\hbar},\frac{\beta}{\hbar}\right)=e^{i\theta_{k,M}}\Omega Z_{0}\mathcal{O}.
\end{equation}
The explicit definitions are
\begin{align}
\Omega & =i^{\frac{1}{2}\left\{ \left(N+M\right)^{2}-N^{2}\right\} -\frac{M}{2}},\nonumber \\
Z_{0} & =k^{-\frac{M}{2}}\prod_{j<j'}^{M}2\sinh\frac{\nu_{N+j}-\nu_{N+j'}}{2k},\nonumber \\
\mathcal{O} & =\frac{\prod_{\overline{n}<\overline{n}'}^{N}2\sinh\frac{\alpha_{\overline{n}}-\alpha_{\overline{n}'}}{2k}\prod_{n}^{N}\prod_{m=n+1}^{N+M}2\sinh\frac{\nu_{n}-\nu_{m}}{2k}}{\prod_{\overline{n}}^{N}\prod_{m}^{N+M}2\cosh\frac{\alpha_{\overline{n}}-\nu_{m}}{2k}}\prod_{n}^{N}\frac{1}{2\cosh\frac{\nu_{n}+i\pi M}{2}}.
\end{align}
$Z_{0}$ is independent of $\alpha_{n}$ and $\beta_{n}$, while $\mathcal{O}$ depends on them. First, by substituting \eqref{eq:RepABJ} into $Z_{0}$, we get
\begin{align}
Z_{0} & =i^{-\frac{1}{2}M\left(M-1\right)}k^{-\frac{M}{2}}\prod_{j<j'}^{M}2\sin\frac{\pi}{k}\left(j'-j\right)\nonumber \\
 & =i^{-\frac{1}{2}M\left(M-1\right)}Z_{k,M}^{\left(\mathrm{CS}\right)},
\end{align}
where $Z_{k,M}^{\left(\mathrm{CS}\right)}$ is defined in \eqref{eq:ZnormDef}. Next, we combine $\Omega$ and the phase $i^{-\frac{1}{2}M\left(M-1\right)}$ appeared in the above expression, and we get
\begin{equation}
i^{-\frac{1}{2}M\left(M-1\right)}\Omega=i^{NM}.
\end{equation}
Finally, we combine this phase factor and $\mathcal{O}$. By substituting \eqref{eq:RepABJ} and using the identity
\begin{equation}
\prod_{j}^{M}f\left(t_{M,j}\right)=\prod_{j}^{M}f\left(-t_{M,j}\right),
\end{equation}
we get
\begin{align}
i^{NM}\mathcal{O} & =i^{NM}\prod_{\overline{n}}^{N}\frac{1}{\prod_{j}^{M}2\cosh\frac{\alpha_{\overline{n}}-2\pi it_{M,j}}{2k}}\nonumber \\
 & \,\,\,\times\frac{\prod_{\overline{n}<\overline{n}'}^{N}2\sinh\frac{\alpha_{\overline{n}}-\alpha_{\overline{n}'}}{2k}\prod_{n<n'}^{N}2\sinh\frac{\beta_{n}-\beta_{n'}}{2k}}{\prod_{\overline{n}}^{N}\prod_{n}^{N}2\cosh\frac{\alpha_{\overline{n}}-\beta_{n}}{2k}}\prod_{n}^{N}\frac{\prod_{j}^{M}2\sinh\frac{\beta_{n}-2\pi it_{M,j}}{2k}}{2\cosh\frac{\beta_{n}+i\pi M}{2}}.
\end{align}
We again use the determinant formula \eqref{eq:CauchyDet}, and then we get
\begin{align}
i^{NM}\mathcal{O} & =\hbar^{N}\det\left(\left[\bra{\alpha_{\overline{n}}}i^{M}\frac{1}{\prod_{j}^{M}2\cosh\frac{\widehat{q}-2\pi it_{M,j}}{2k}}\frac{1}{2\cosh\frac{\widehat{p}}{2}}\frac{\prod_{j}^{M}2\sinh\frac{\widehat{q}-2\pi it_{M,j}}{2k}}{2\cosh\frac{\widehat{q}+i\pi M}{2}}\ket{\beta_{n}}\right]_{\overline{n},n}^{N\times N}\right).
\end{align}
Finally, by combining all of the above results, we obtain
\begin{equation}
\widetilde{Z}_{k,M}^{\left(+1,-1\right)}\left(N;\frac{\alpha}{\hbar},\frac{\beta}{\hbar}\right)=e^{i\theta_{k,M}}Z_{k,M}^{\left(\mathrm{CS}\right)}\hbar^{N}\det\left(\left[\bra{\alpha_{\overline{n}}}\widehat{\rho}_{M}^{\left(+1,-1\right)}\ket{\beta_{n}}\right]_{\overline{n},n}^{N\times N}\right),\label{eq:ZABJ}
\end{equation}
where $\widehat{\rho}_{M}^{\left(+1,-1\right)}=\widehat{\rho}_{M}^{\mathrm{ABJ}}$ defined in \eqref{eq:rhoABJ}. This form is clearly \eqref{eq:ZtildeGenRes}. Therefore we have accomplished applying the Fermi gas formalism to the ABJ matrix model.

\section{Fermi gas formalism for general ranks\label{sec:FGF}}

As explained in section \ref{subsec:Strategy}, what we will achieve in this paper is to transform the deformed factors to the form of \eqref{eq:ZtildeGenRes}, and in this section we carry out this plan. In section \ref{subsec:Target}, we present deformed factors we focus on in this paper and the result of the Fermi gas approach. In section \ref{subsec:Calculation}, we show its derivation. In section \ref{subsec:HWtrans}, we prove that the Hanany-Witten transition holds on the level of deformed factors. 

\subsection{Our setup and results\label{subsec:Target}}

\begin{figure}
\begin{centering}
\includegraphics[scale=0.8]{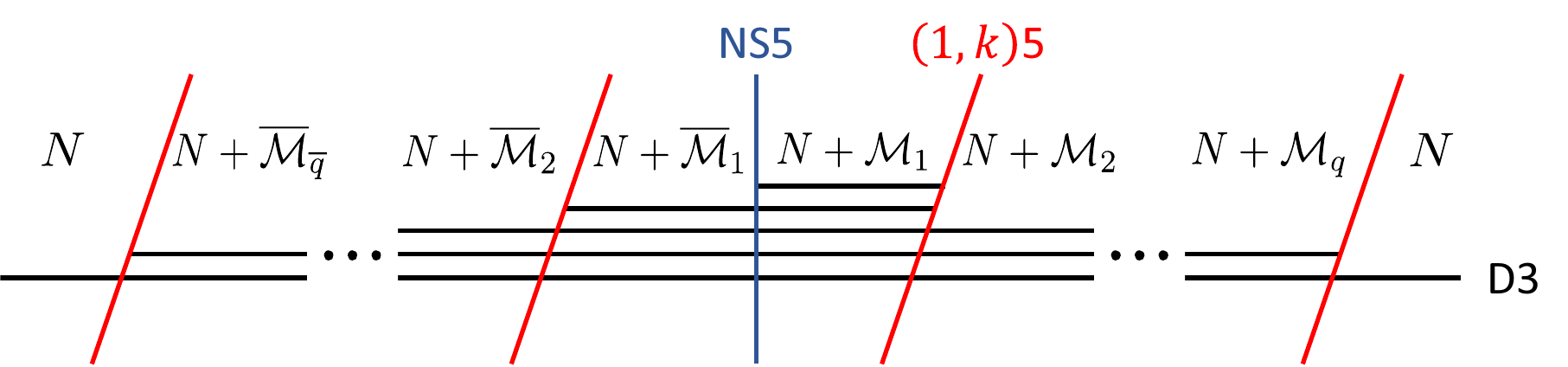}
\par\end{centering}
\caption{\label{fig:BraneConfig}The brane configurations we consider in this paper. The blue line and the red lines represent an NS5-brane and $\left(1,k\right)$5-branes, respectively. The black lines represent coincident D3-branes, and each symbol above the D3-branes refers to the number of D3-branes on each segment.}
\end{figure}
We first explain the brane configurations in type IIB string theory which we focus on in this paper (figure \ref{fig:BraneConfig}). These brane configurations involve an NS5-brane at the center, any number of $\left(1,k\right)$5-branes on both sides of the NS5-brane and D3-branes stretched between two 5-branes. We parameterize these brane configurations as
\begin{equation}
\left\langle 0\circ\overline{\mathcal{M}}_{\overline{q}}\circ\ldots\circ\overline{\mathcal{M}}_{1}\bullet\mathcal{M}_{1}\circ\ldots\circ\mathcal{M}_{q}\circ0\right\rangle .\label{eq:ConfiguGen}
\end{equation}
We restrict the number of D3-branes stretched between an NS5-brane and a $\left(1,k\right)$5-brane to equal to or less than $k$ as is the case with ABJ theory. We also assume that the number of D3-branes monotonically increases with approaching the NS5-brane at the center, and the growth of the number of the D3-branes near the NS5-brane is almost larger than that far from the NS5-brane. To give a rigorous description of the above statement, we define the difference between the adjacent ranks:
\begin{equation}
M_{a}=\mathcal{M}_{a}-\mathcal{M}_{a+1},\quad\overline{M}_{\overline{a}}=\mathcal{M}_{\overline{a}}-\mathcal{M}_{\overline{a}+1},\label{eq:McalToM}
\end{equation}
where $\mathcal{M}_{q+1}=\overline{\mathcal{M}}_{\overline{q}+1}=0$. These values satisfy the following restrictions:
\begin{align}
 & 0\leq M_{a}\leq k,\quad0\leq\overline{M}_{\overline{a}}\leq k\quad\left(\text{for all }a\text{ and }\overline{a}\right),\nonumber \\
 & M_{a}+1\geq M_{b},\quad\overline{M}_{\overline{a}}+1\geq\overline{M}_{\overline{b}}\quad\left(\text{for all }a<b\text{ and }\overline{a}<\overline{b}\right),\nonumber \\
 & M_{a}+\overline{M}_{\overline{a}}\leq k+1\quad\left(\text{for all }a\text{ and }\overline{a}\right).\label{eq:ConfigCond}
\end{align}
We explain the physical meaning of these restrictions in the second and the third lines at the end of this section. Note that these brane configurations include the brane configuration appeared in section \ref{subsec:ABJcase} as a special case.

The deformed factors \eqref{eq:PartGenDef} corresponding to the part of brane configurations \eqref{eq:ConfiguGen} are
\begin{align}
 & Z_{k,\boldsymbol{M}}^{\left(-\overline{q},+1,-q\right)}\left(N;\frac{\alpha}{\hbar},\frac{\beta}{\hbar}\right)\nonumber \\
 & =\frac{i^{-\frac{1}{2}\left\{ \left(N+\overline{\mathcal{M}}_{1}\right)^{2}-N^{2}\right\} +\frac{1}{2}\left\{ \left(N+\mathcal{M}_{1}\right)^{2}-N^{2}\right\} }}{\prod_{\overline{a}}^{\overline{q}}\left(N+\overline{\mathcal{M}}_{\overline{a}}\right)!\prod_{a}^{q}\left(N+\mathcal{M}_{a}\right)!}\int\left(\prod_{\overline{a}}^{\overline{q}}\frac{d^{N+\overline{\mathcal{M}}_{\overline{a}}}\mu^{\left(\overline{a}\right)}}{\hbar^{N+\overline{\mathcal{M}}_{\overline{a}}}}\right)\left(\prod_{a}^{q}\frac{d^{N+\mathcal{M}_{a}}\nu^{\left(a\right)}}{\hbar^{N+\mathcal{M}_{a}}}\right)\nonumber \\
 & \,\,\,\times e^{S_{k}\left(N+\overline{\mathcal{M}}_{1};\frac{\mu^{\left(1\right)}}{\hbar}\right)}e^{S_{-k}\left(N+\mathcal{M}_{1};\frac{\nu^{\left(1\right)}}{\hbar}\right)}\prod_{\overline{a}}^{\overline{q}}Z\left(N+\overline{\mathcal{M}}_{\overline{a}+1},N+\overline{\mathcal{M}}_{\overline{a}};\frac{\mu^{\left(\overline{a}+1\right)}}{\hbar},\frac{\mu^{\left(\overline{a}\right)}}{\hbar}\right)\nonumber \\
 & \,\,\,\times Z\left(N+\overline{\mathcal{M}}_{1},N+\mathcal{M}_{1};\frac{\mu^{\left(1\right)}}{\hbar},\frac{\nu^{\left(1\right)}}{\hbar}\right)\prod_{a}^{q}Z\left(N+\mathcal{M}_{a},N+\mathcal{M}_{a+1};\frac{\nu^{\left(a\right)}}{\hbar},\frac{\nu^{\left(a+1\right)}}{\hbar}\right),\label{eq:ZpartGen}
\end{align}
where $\mathcal{M}_{q+1}=\overline{\mathcal{M}}_{\overline{q}+1}=0$,
\begin{equation}
\boldsymbol{M}=\left(\overline{\mathcal{M}}_{\overline{q}},\ldots,\overline{\mathcal{M}}_{1},\mathcal{M}_{1},\ldots,\mathcal{M}_{q}\right),\label{eq:MvecDef}
\end{equation}
and we introduced the following notation:
\begin{equation}
\left(-\overline{q},+1,-q\right)=(\overset{\overline{q}}{\overbrace{-1,\ldots,-1}},+1,\overset{q}{\overbrace{-1,\ldots,-1}}).
\end{equation}
Note that the order of the production $\prod_{\overline{a}}^{\overline{q}}$ is $\overline{q}=\overline{a},\overline{a}-1,\ldots,1$ in spirit though we do not mind this point hereafter.

We carry out the computation of the deformed factors in the next section. Before that, we devote the rest of this section to present the result and discuss its physical meaning. We transform $Z_{k,\boldsymbol{M}}^{\left(-\overline{q},+1,-q\right)}$ according to the rule \eqref{eq:BraketChange}, and the result is denoted by $\widetilde{Z}_{k,\boldsymbol{M}}^{\left(-\overline{q},+1,-q\right)}$ as explained in section \ref{subsec:Strategy}. We find that
\begin{align}
\widetilde{Z}_{k,\boldsymbol{M}}^{\left(-\overline{q},+1,-q\right)}\left(N;\frac{\alpha}{\hbar},\frac{\beta}{\hbar}\right) & =e^{i\Theta_{k,\boldsymbol{M}}^{\left(-\overline{q},+1,-q\right)}}\left(\prod_{\overline{a}}^{\overline{q}}Z_{k,\overline{M}_{\overline{a}}}^{\left(\mathrm{CS}\right)}\right)\left(\prod_{a}^{q}Z_{k,M_{a}}^{\left(\mathrm{CS}\right)}\right)\nonumber \\
 & \,\,\,\times\left(\prod_{\overline{a}}^{\overline{q}-1}\prod_{\overline{b}=\overline{a}+1}^{\overline{q}}Z_{k,\overline{M}_{\overline{a}},\overline{M}_{\overline{b}}}^{\left(\Iside\right)}\right)\left(\prod_{\overline{a}}^{\overline{q}}\prod_{a}^{q}Z_{k,\overline{M}_{\overline{a}},M_{a}}^{\left(\Icross\right)}\right)\left(\prod_{a}^{q-1}\prod_{b=a+1}^{q}Z_{k,M_{a},M_{b}}^{\left(\Iside\right)}\right)\nonumber \\
 & \,\,\,\times\hbar^{N}\det\left(\left[\bra{\alpha_{\overline{n}}}\widehat{\rho}_{\boldsymbol{M}}^{\left(-\overline{q},+1,-q\right)}\ket{\beta_{n}}\right]_{\overline{n},n}^{N\times N}\right).\label{eq:ZpGenRes}
\end{align}
The definition of each symbol is as follows. The phase symbol is
\begin{equation}
\Theta_{k,\boldsymbol{M}}^{\left(-\overline{q},+1,-q\right)}=\frac{1}{2}\left(-\sum_{\overline{a}}^{\overline{q}}\theta_{k,\overline{M}_{\overline{a}}}+\theta_{k,\mathcal{M}_{1}-\overline{\mathcal{M}}_{1}}+\sum_{a}^{q}\theta_{k,M_{a}}\right),\label{eq:PhaseAllDef}
\end{equation}
where $\theta_{k,M}$ appeared in the ABJ theory and is defined in \eqref{eq:phaseABJ}. The other $N$-independent factors are
\begin{align}
Z_{k,M}^{\left(\mathrm{CS}\right)} & =\frac{1}{k^{\frac{M}{2}}}\prod_{j<j'}^{M}2\sin\frac{\pi}{k}\left(j'-j\right),\nonumber \\
Z_{k,M,M'}^{\left(\Iside\right)} & =\begin{cases}
\frac{i^{\left(M-1\right)M'}}{k^{M'}}\prod_{j'}^{M'}\prod_{t_{M,j}\neq t_{M',j'}}^{M}2\sin\frac{\pi}{k}\left(t_{M',j'}-t_{M,j}\right) & \left(M+M'=\text{even}\right)\\
\frac{i^{MM'}}{2^{M'}}\prod_{j'}^{M'}\prod_{j}^{M}2\sin\frac{\pi}{k}\left(t_{M',j'}-t_{M,j}\right) & \left(M+M'=\text{odd}\right)
\end{cases},\nonumber \\
Z_{k,\overline{M},M}^{\left(\Icross\right)} & =\prod_{\overline{j}}^{\overline{M}}\prod_{j}^{M}\frac{1}{2\cos\frac{\pi}{k}\left(t_{\overline{M},\overline{j}}-t_{M,j}\right)},\label{eq:ZnormDef}
\end{align}
where $\prod_{t_{M,j}\neq t_{M',j'}}^{M}$ means that $j$ runs from $1$ to $M$ except for the case when $t_{M,j}=t_{M',j'}$. $Z_{k,M,M'}^{\left(\Iside\right)}$ comes from 1-loop contributions of vector multiplets \eqref{eq:VectorOneLoop} , while $Z_{k,\overline{M},M}^{\left(\Icross\right)}$ comes from 1-loop contributions of hypermultiplets \eqref{eq:HyperOneLoop} as we will show in the next section. These factors are positive as can be seen from this definition. The density matrix is
\begin{align}
\widehat{\rho}_{\boldsymbol{M}}^{\left(-\overline{q},+1,-q\right)} & =i^{\mathcal{M}_{1}-\overline{\mathcal{M}}_{1}}\left(\prod_{\overline{a}}^{\overline{q}}\frac{\prod_{\overline{j}_{\overline{a}}}^{\overline{M}_{\overline{a}}}2\sinh\frac{\widehat{q}-2\pi it_{\overline{M}_{\overline{a}},\overline{j}_{\overline{a}}}}{2k}}{2\cosh\frac{\widehat{q}-i\pi\overline{M}_{\overline{a}}}{2}}\right)\left(\prod_{a}^{q}\frac{1}{\prod_{j_{a}}^{M_{a}}2\cosh\frac{\widehat{q}-2\pi it_{M_{a},j_{a}}}{2k}}\right)\nonumber \\
 & \,\,\,\times\frac{1}{2\cosh\frac{\widehat{p}}{2}}\left(\prod_{a}^{q}\frac{\prod_{j_{a}}^{M_{a}}2\sinh\frac{\widehat{q}-2\pi it_{M_{a},j_{a}}}{2k}}{2\cosh\frac{\widehat{q}+i\pi M_{a}}{2}}\right)\left(\prod_{\overline{a}}^{\overline{q}}\frac{1}{\prod_{\overline{j}_{\overline{a}}}^{\overline{M}_{\overline{a}}}2\cosh\frac{\widehat{q}-2\pi it_{\overline{M}_{\overline{a}},\overline{j}_{\overline{a}}}}{2k}}\right).\label{eq:rhoGen}
\end{align}
Note that $\widetilde{Z}_{k,\boldsymbol{M}}^{\left(-\overline{q},+1,-q\right)}$ reduces to \eqref{eq:ZABJ} when $\overline{q}=0$ and $q=1$.

We give some comments on the setup and the results. First, the role of $s=\pm1$ seems to be asymmetric in our setup \eqref{eq:ConfiguGen}. However, it is in fact symmetric in the following meaning. If we take complex conjugate for the whole matrix models \eqref{eq:PFall}, we get matrix models whose $s=\pm1$ are exchanged. Hence the computability is not changed under this exchange. Therefore, the brane configurations which can be applied to the Fermi gas formalism by using our method are the combination of \eqref{eq:ConfiguTriv} and \eqref{eq:ConfiguGen} and also the combination of \eqref{eq:ConfiguTriv} and $s=\pm1$ exchanged \eqref{eq:ConfiguGen}.

Second, the result \eqref{eq:ZpGenRes} is also valid for $N=0$, in which case the last line of \eqref{eq:ZpGenRes}, namely the Fermi gas factor vanishes. These are the matrix models computing the partition functions of $\mathcal{N}=4$ superconformal Chern-Simons theories described by linear quiver diagrams \cite{Gaiotto:2008sd,Hosomichi:2008jd,Imamura:2008dt}, because $Z_{k,\boldsymbol{M}}^{\left(-\overline{q},+1,-q\right)}$ are matrix models of these theories as explained below \eqref{eq:PartGenDef} and in this case the following identity holds:
\begin{equation}
Z_{k,\boldsymbol{M}}^{\left(-\overline{q},+1,-q\right)}\left(N=0\right)=\widetilde{Z}_{k,\boldsymbol{M}}^{\left(-\overline{q},+1,-q\right)}\left(N=0\right).
\end{equation}

Third, we comment on the positive $N$-independent factor $Z_{k,M}^{\left(\mathrm{CS}\right)}$. It is clear by comparing the brane configurations \eqref{eq:ConfiguGen} and the result \eqref{eq:ZpGenRes} that $Z_{k,M}^{\left(\mathrm{CS}\right)}$ appears for each stack of coincident D3-branes stretched between the same NS5-brane and the same $\left(1,k\right)$5-brane. This is the natural result because it is known that the worldvolume theory of this brane configuration (namely, a stack of $M$ coincident D3-branes stretched between an NS5-brane and a $\left(1,k\right)$5-brane) is $\mathrm{U}\left(M\right)_{k}$ pure Chern-Simons theory \cite{Kitao:1998mf,Bergman:1999na} and $Z_{k,M}^{\left(\mathrm{CS}\right)}$ is the partition function of the same theory on $S^{3}$ \cite{Marino:2002fk,Marino:2004uf}.

Fourth, we comment on the other positive $N$-independent factors $Z_{k,M,M'}^{\left(\Iside\right)}$ and $Z_{k,\overline{M},M}^{\left(\Icross\right)}$. First, $Z_{k,M,M'}^{\left(\Iside\right)}$ appears through \eqref{eq:ZintForm}, and the left-hand side of \eqref{eq:ZintForm} diverges when the parameters exceed the second line of the restriction \eqref{eq:ConfigCond}, namely when $M+2=M'$ (see also the argument in the below \eqref{eq:ZintForm}). $Z_{k,\overline{M},M}^{\left(\Icross\right)}$ also diverges when the parameters exceed the third line of the restriction \eqref{eq:ConfigCond}, namely when $M+\overline{M}=k+2$. These divergences do not disappear by the procedure in footnote \ref{fn:Divergence}. It was argued in \cite{Kapustin:2010mh} that when a matrix model diverges even after this procedure, the corresponding gauge theory is ``bad'' in the context of \cite{Gaiotto:2008ak}. Furthermore, the more explicit suggestion appeared in \cite{Nosaka:2017ohr,Nosaka:2018eip} for the worldvolume theory of the brane configuration $\left\langle 0\bullet\overline{M}\circ M\bullet0\right\rangle _{0}$ (the corresponding matrix model is the complex conjugate of one of $\left\langle 0\bullet\overline{M}\circ M\bullet0\right\rangle _{0}$, which is the special case of \eqref{eq:ConfiguGen}, and especially the positive factors are equal). $Z_{k,\overline{M},M}^{\left(\Icross\right)}$ appears in this case, and they revealed that the gauge theory is ``good'', ``ugly'' or ``bad'' when the parameters satisfy $M+\overline{M}\leq k$, $M+\overline{M}=k+1$ or $M+\overline{M}\geq k+2$, respectively. Furthermore, the restrictions in the second and the third lines of \eqref{eq:ConfigCond} are equivalent under the Hanany-Witten transition as we will explain in section \ref{subsec:HWtrans}, and thus it is natural to expect that the physical origin of divergence of $Z_{k,M,M'}^{\left(\Iside\right)}$ and $Z_{k,\overline{M},M}^{\left(\Icross\right)}$ are the same. Therefore, we expect that if any one of the restrictions in the second and the third lines of \eqref{eq:ConfigCond} are not satisfied, the corresponding gauge theory is ``bad''.

Fifth, the Hermitian conjugate of a density matrix associated with a brane configuration \eqref{eq:ConfiguGen} is one associated with the inverse of the brane configuration, namely
\begin{equation}
\left\langle 0\circ\mathcal{M}_{q}\circ\ldots\circ\mathcal{M}_{1}\bullet\overline{\mathcal{M}}_{1}\circ\ldots\circ\overline{\mathcal{M}}_{\overline{q}}\circ0\right\rangle .
\end{equation}
Explicitly, the identity
\begin{equation}
\left(\widehat{\rho}_{\boldsymbol{M}}^{\left(-\overline{q},+1,-q\right)}\right)^{\dagger}=\widehat{\rho}_{\boldsymbol{M}^{t}}^{\left(-q,+1,-\overline{q}\right)},\label{eq:DensHermit}
\end{equation}
holds, where
\begin{equation}
\boldsymbol{M}^{t}=\left(\mathcal{M}_{q},\ldots,\mathcal{M}_{1},\overline{\mathcal{M}}_{1},\ldots,\overline{\mathcal{M}}_{\overline{q}}\right).
\end{equation}
Note that this identity comes from the identity
\begin{equation}
\left\{ Z_{k,\boldsymbol{M}}^{\left(-\overline{q},+1,-q\right)}\left(N;\frac{\alpha}{\hbar},\frac{\beta}{\hbar}\right)\right\} ^{*}=Z_{k,\boldsymbol{M}^{t}}^{\left(-q,+1,-\overline{q}\right)}\left(N;\frac{\beta}{\hbar},\frac{\alpha}{\hbar}\right).
\end{equation}

\subsection{Derivation\label{subsec:Calculation}}

In this section we compute the general deformed factors \eqref{eq:ZpartGen} and transform them into \eqref{eq:ZpGenRes}. The general flow of computation is similar to the ABJ case in section \ref{subsec:ABJcase}, and thus we omit similar points accordingly.

We start with using the determinant formula \eqref{eq:CauchyDet} to \eqref{eq:ZpartGen}:
\begin{align}
 & Z_{k,\boldsymbol{M}}^{\left(-\overline{q},+1,-q\right)}\left(N;\frac{\alpha}{\hbar},\frac{\beta}{\hbar}\right)\nonumber \\
 & =\frac{i^{-\frac{1}{2}\left\{ \left(N+\overline{\mathcal{M}}_{1}\right)^{2}-N^{2}\right\} +\frac{1}{2}\left\{ \left(N+\mathcal{M}_{1}\right)^{2}-N^{2}\right\} }}{\prod_{\overline{a}}^{\overline{q}}\left(N+\overline{\mathcal{M}}_{\overline{a}}\right)!\prod_{a}^{q}\left(N+\mathcal{M}_{a}\right)!}\int\left(\prod_{\overline{a}}^{\overline{q}}\frac{d^{N+\overline{\mathcal{M}}_{\overline{a}}}\mu^{\left(\overline{a}\right)}}{\hbar^{N+\overline{\mathcal{M}}_{\overline{a}}}}\right)\left(\prod_{a}^{q}\frac{d^{N+\mathcal{M}_{a}}\nu^{\left(a\right)}}{\hbar^{N+\mathcal{M}_{a}}}\right)\nonumber \\
 & \,\,\,\times e^{\frac{i}{2\hbar}\sum_{\overline{m}_{1}}^{N+\overline{\mathcal{M}}_{1}}\left(\mu_{\overline{m}_{1}}^{\left(1\right)}\right)^{2}}e^{-\frac{i}{2\hbar}\sum_{m_{1}}^{N+\mathcal{M}_{1}}\left(\nu_{m_{1}}^{\left(1\right)}\right)^{2}}\nonumber \\
 & \,\,\,\times\prod_{\overline{a}}^{\overline{q}}\det\left(\begin{array}{c}
\left[\hbar\bra{\mu_{\overline{m}_{\overline{a}+1}}^{\left(\overline{a}+1\right)}}\frac{1}{2\cosh\frac{\widehat{p}-i\pi\overline{M}_{\overline{a}}}{2}}\ket{\mu_{\overline{m}_{\overline{a}}}^{\left(\overline{a}\right)}}\right]_{\overline{m}_{\overline{a}+1},\overline{m}_{\overline{a}}}^{\left(N+\overline{\mathcal{M}}_{\overline{a}+1}\right)\times\left(N+\overline{\mathcal{M}}_{\overline{a}}\right)}\\
\left[\frac{\hbar}{\sqrt{k}}\bbraket{2\pi it_{\overline{M}_{\overline{a}},\overline{j}_{\overline{a}}}|\mu_{\overline{m}_{\overline{a}}}^{\left(\overline{a}\right)}}\right]_{\overline{j}_{\overline{a}},\overline{m}_{\overline{a}}}^{\overline{M}_{\overline{a}}\times\left(N+\overline{\mathcal{M}}_{\overline{a}}\right)}
\end{array}\right)\nonumber \\
 & \,\,\,\times Z\left(N+\overline{\mathcal{M}}_{1},N+\mathcal{M}_{1};\frac{\mu^{\left(1\right)}}{\hbar},\frac{\nu^{\left(1\right)}}{\hbar}\right)\nonumber \\
 & \,\,\,\times\prod_{a}^{q}\det\left(\begin{array}{cc}
\left[\hbar\bra{\nu_{m_{a}}^{\left(a\right)}}\frac{1}{2\cosh\frac{\widehat{p}+i\pi M_{a}}{2}}\ket{\nu_{m_{a+1}}^{\left(a+1\right)}}\right]_{m_{a},m_{a+1}}^{_{\times\left(N+\mathcal{M}_{a+1}\right)}^{\left(N+\mathcal{M}_{a}\right)}} & \left[\frac{\hbar}{\sqrt{k}}\brakket{\nu_{m_{a}}^{\left(a\right)}|-2\pi it_{M_{a},j_{a}}}\right]_{m_{a},j_{a}}^{_{\times M_{a}}^{\left(N+\mathcal{M}_{a}\right)}}\end{array}\right).
\end{align}
It should be emphasized that we have applied \eqref{eq:CauchyDet} also to $Z\left(N+\overline{\mathcal{M}}_{1},N+\mathcal{M}_{1}\right)$. As explained in section \ref{subsec:Strategy}, we eliminate all the Fresnel factors and transform $Z_{k,\boldsymbol{M}}^{\left(-\overline{q},+1,-q\right)}$ into $\widetilde{Z}_{k,\boldsymbol{M}}^{\left(-\overline{q},+1,-q\right)}$ according to \eqref{eq:BraketChange}. We further proceed with the computation in the same manner as section \ref{subsec:ABJcase}. More explicitly, we first use the formulas \eqref{eq:Simptoq} and \eqref{eq:Vecptoq} so that all momentum operators and momentum eigenvectors except for those in $Z\left(N+\overline{\mathcal{M}}_{1},N+\mathcal{M}_{1}\right)$ become position operators and position eigenvectors respectively. Second, we diagonalize all determinants except for $Z\left(N+\overline{\mathcal{M}}_{1},N+\mathcal{M}_{1}\right)$ using the formula \eqref{eq:Diagonalize}. Finally, we restore $Z\left(N+\overline{\mathcal{M}}_{1},N+\mathcal{M}_{1}\right)$ using the determinant formula \eqref{eq:CauchyDet} backward. After these computations we obtain
\begin{align}
 & \widetilde{Z}_{k,\boldsymbol{M}}^{\left(-\overline{q},+1,-q\right)}\left(N;\frac{\alpha}{\hbar},\frac{\beta}{\hbar}\right)\nonumber \\
 & =i^{-\frac{1}{2}\left\{ \left(N+\overline{\mathcal{M}}_{1}\right)^{2}-N^{2}\right\} +\frac{1}{2}\left\{ \left(N+\mathcal{M}_{1}\right)^{2}-N^{2}\right\} }i^{\frac{\overline{\mathcal{M}}_{1}}{2}-\frac{\mathcal{M}_{1}}{2}}e^{i\Theta_{k,\boldsymbol{M}}^{\left(-\overline{q},+1,-q\right)}}k^{-\frac{\overline{\mathcal{M}}_{1}}{2}-\frac{\mathcal{M}_{1}}{2}}\nonumber \\
 & \,\,\,\times\int\left(\prod_{\overline{a}}^{\overline{q}}d^{N+\overline{\mathcal{M}}_{\overline{a}}}\mu^{\left(\overline{a}\right)}\right)\left(\prod_{a}^{q}d^{N+\mathcal{M}_{a}}\nu^{\left(a\right)}\right)\nonumber \\
 & \,\,\,\times\prod_{\overline{a}}^{\overline{q}}\left\{ \left(\prod_{\overline{m}_{\overline{a}+1}}^{N+\overline{\mathcal{M}}_{\overline{a}+1}}\bra{\mu_{\overline{m}_{\overline{a}+1}}^{\left(\overline{a}+1\right)}}\frac{1}{2\cosh\frac{\widehat{q}-i\pi\overline{M}_{\overline{a}}}{2}}\ket{\mu_{\overline{m}_{\overline{a}+1}}^{\left(\overline{a}\right)}}\right)\left(\prod_{\overline{j}_{\overline{a}}}^{\overline{M}_{\overline{a}}}\braket{2\pi it_{\overline{M}_{\overline{a}},\overline{j}_{\overline{a}}}|\mu_{N+\overline{\mathcal{M}}_{\overline{a}+1}+\overline{j}_{\overline{a}}}^{\left(\overline{a}\right)}}\right)\right\} \nonumber \\
 & \,\,\,\times\frac{\prod_{\overline{m}_{1}<\overline{m}_{1}'}^{N+\overline{\mathcal{M}}_{1}}2\sinh\frac{\mu_{\overline{m}_{1}}^{\left(1\right)}-\mu_{\overline{m}_{1}'}^{\left(1\right)}}{2k}\prod_{m_{1}<m_{1}'}^{N+\mathcal{M}_{1}}2\sinh\frac{\nu_{m_{1}}^{\left(1\right)}-\nu_{m_{1}'}^{\left(1\right)}}{2k}}{\prod_{\overline{m}_{1}}^{N+\overline{\mathcal{M}}_{1}}\prod_{m_{1}}^{N+\mathcal{M}_{1}}2\cosh\frac{\mu_{\overline{m}_{1}}^{\left(1\right)}-\nu_{m_{1}}^{\left(1\right)}}{2k}}\nonumber \\
 & \,\,\,\times\prod_{a}^{q}\left\{ \left(\prod_{m_{a+1}}^{N+\mathcal{M}_{a+1}}\bra{\nu_{m_{a}}^{\left(a\right)}}\frac{1}{2\cosh\frac{\widehat{q}+i\pi M_{a}}{2}}\ket{\nu_{m_{a+1}}^{\left(a+1\right)}}\right)\left(\prod_{j_{a}}^{M_{a}}\braket{\nu_{N+\mathcal{M}_{a+1}+j_{a}}^{\left(a\right)}|-2\pi it_{M_{a},j_{a}}}\right)\right\} .\label{eq:ZpartTrans1}
\end{align}
All the integrations seem to be performed using the delta  functions coming from inner products of position eigenvectors. However, fake divergences appear with a naive computation.\footnote{This divergence already appeared in \cite{Moriyama:2017gye}. The reason this divergence appear is we used the determinant formula \eqref{eq:CauchyDet}.} Therefore, we introduce small convergence factors into all constant position eigenvectors:
\begin{align}
\bra{2\pi it_{\overline{M}_{\overline{a}},\overline{j}_{\overline{a}}}} & =\lim_{\overline{\epsilon}_{\overline{a}}\rightarrow0}\bra{2\pi it_{\overline{M}_{\overline{a}},\overline{j}_{\overline{a}}}+\overline{\epsilon}_{\overline{a}}},\nonumber \\
\ket{-2\pi it_{M_{a},j_{a}}} & =\lim_{\epsilon_{a}\rightarrow0}\ket{-2\pi it_{M_{a},j_{a}}+\epsilon_{a}}.
\end{align}

We have to consider poles to perform the integration because of the argument in appendix \ref{sec:integral} as with the ABJ theory. To come to the point, there are no poles in the region where integration contour passes due to the restriction \eqref{eq:ConfigCond}, so that we can perform all the integration using the property of the delta functions as with the ABJ theory. First, we perform the integration over $\nu_{m_{1}}^{\left(1\right)}$. We should be careful about the case when $m_{1}=N+\mathcal{M}_{2}+j_{1}$ ($1\leq j_{1}\leq M_{1}$). The part of the function of $\nu_{N+\mathcal{M}_{2}+j_{1}}^{\left(1\right)}$ having poles and the region where this part has no poles is
\begin{equation}
\frac{1}{\prod_{\overline{m}_{1}}^{N+\overline{\mathcal{M}}_{1}}2\cosh\frac{\mu_{\overline{m}_{1}}^{\left(1\right)}-\nu_{m_{1}}^{\left(1\right)}}{2k}},\quad\text{No pole region: }\left|\mathrm{Im}\left(\nu_{m_{1}}^{\left(1\right)}\right)\right|<\pi k.
\end{equation}
Moreover, the restriction in the first line of \eqref{eq:ConfigCond} leads $\left|2\pi t_{M_{1},j_{1}}\right|\leq\pi\left(k-1\right)$, so that integration contour passes through no poles. After performing the integration over $\nu_{m_{1}}^{\left(1\right)}$, we perform the integration over $\nu_{m_{2}}^{\left(2\right)}$. In this case, there are two types of functions of $\nu_{N+\mathcal{M}_{3}+j_{2}}^{\left(2\right)}$ ($1\leq j_{2}\leq M_{2}$) having poles:
\begin{align}
\frac{1}{\prod_{\overline{m}_{1}}^{N+\overline{\mathcal{M}}_{1}}2\cosh\frac{\mu_{\overline{m}_{1}}^{\left(1\right)}-\nu_{m_{2}}^{\left(2\right)}}{2k}}, & \quad\text{No pole region: }\left|\mathrm{Im}\left(\nu_{m_{2}}^{\left(2\right)}\right)\right|<\pi k,\nonumber \\
\frac{\prod_{j_{1}}^{M_{1}}2\sinh\frac{\nu_{m_{2}}^{\left(2\right)}+2\pi it_{M_{1},j_{1}}}{2k}}{2\cosh\frac{\nu_{m_{2}}^{\left(2\right)}+i\pi M_{1}}{2}}, & \quad\text{No pole region: }\left|\mathrm{Im}\left(\nu_{m_{2}}^{\left(2\right)}\right)\right|<\pi\left(M_{1}+1\right).
\end{align}
Moreover, the restriction in the first and the second lines of \eqref{eq:ConfigCond} leads $\left|2\pi t_{M_{2},j_{2}}\right|\leq\pi\left(k-1\right)$ and $\left|2\pi t_{M_{2},j_{2}}\right|\leq\pi M_{1}$ respectively, so that integration contour again passes through no poles. The similar argument holds for the other $\nu_{m_{a}}^{\left(a\right)}$. Therefore, we can perform the integration over $\nu_{m_{a}}^{\left(a\right)}$ in the order $a=1,2,\ldots,q$. We can also carry out the integration over $\mu_{\overline{m}_{\overline{a}}}^{\left(\overline{a}\right)}$ in the order $\overline{a}=1,2,\ldots,\overline{q}$ in the similar manner. The only difference is that there are factors
\begin{equation}
\frac{1}{\prod_{m_{1}}^{N+\mathcal{M}_{1}}2\cosh\frac{\mu_{\overline{m}_{1}}^{\left(1\right)}-\nu_{m_{1}}^{\left(1\right)}}{2k}},
\end{equation}
in \eqref{eq:ZpartTrans1} which now contain imaginary constant $-2\pi it_{M_{a},j_{a}}$ instead of real number $\nu_{m_{1}}^{\left(1\right)}$. The region where this function has no poles is
\begin{equation}
\prod_{a}^{q}\frac{1}{\prod_{j_{a}}^{M_{a}}2\cosh\frac{\mu_{\overline{m}_{\overline{a}}}^{\left(\overline{a}\right)}+2\pi it_{M_{a},j_{a}}}{2k}},\quad\text{No pole region: }\left|\mathrm{Im}\left(\mu_{\overline{m}_{\overline{a}}}^{\left(\overline{a}\right)}\right)\right|<\pi\left(k+1-M_{a}\right).
\end{equation}
Moreover, the restriction in the third line of \eqref{eq:ConfigCond} leads $\left|2\pi t_{\overline{M}_{\overline{a}},\overline{j}_{\overline{a}}}\right|\leq\pi\left(k-M_{a}\right)$, so that we can perform the integration over $\mu_{\overline{m}_{\overline{a}}}^{\left(\overline{a}\right)}$. From the above, we can carry out all the integration:
\begin{align}
 & \widetilde{Z}_{k,\bm{M}}^{\left(-\overline{q},+1,-q\right)}\left(N;\frac{\alpha}{\hbar},\frac{\beta}{\hbar}\right)\nonumber \\
 & =i^{-\frac{1}{2}\left\{ \left(N+\overline{\mathcal{M}}_{1}\right)^{2}-N^{2}\right\} +\frac{1}{2}\left\{ \left(N+\mathcal{M}_{1}\right)^{2}-N^{2}\right\} }i^{\frac{\overline{\mathcal{M}}_{1}}{2}-\frac{\mathcal{M}_{1}}{2}}e^{i\Theta_{k,\bm{M}}^{\left(-\overline{q},+1,-q\right)}}k^{-\frac{\overline{\mathcal{M}}_{1}}{2}-\frac{\mathcal{M}_{1}}{2}}\nonumber \\
 & \,\,\,\times\left(\prod_{\overline{a}}^{\overline{q}}\lim_{\overline{\epsilon}_{\overline{a}}\rightarrow0}\right)\left(\prod_{a}^{q}\lim_{\epsilon_{a}\rightarrow0}\right)\prod_{\overline{a}}^{\overline{q}}\prod_{\overline{m}_{\overline{a}+1}}^{N+\overline{\mathcal{M}}_{\overline{a}+1}}\frac{1}{2\cosh\frac{\mu_{\overline{m}_{\overline{a}+1}}-i\pi\overline{M}_{\overline{a}}}{2}}\nonumber \\
 & \,\,\,\times\frac{\prod_{\overline{m}<\overline{m}'}^{N+\overline{\mathcal{M}}_{1}}2\sinh\frac{\mu_{\overline{m}}-\mu_{\overline{m}'}}{2k}\prod_{m<m'}^{N+\mathcal{M}_{1}}2\sinh\frac{\nu_{m}-\nu_{m'}}{2k}}{\prod_{\overline{m}}^{N+\overline{\mathcal{M}}_{1}}\prod_{m}^{N+\mathcal{M}_{1}}2\cosh\frac{\mu_{\overline{m}}-\nu_{m}}{2k}}\prod_{a}^{q}\prod_{m_{a+1}}^{N+\mathcal{M}_{a+1}}\frac{1}{2\cosh\frac{\nu_{m_{a+1}}+i\pi M_{a}}{2}},
\end{align}
where
\begin{equation}
\mu_{\overline{m}}=\begin{cases}
\alpha_{\overline{m}} & \left(1\leq\overline{m}\leq N\right)\\
2\pi it_{\overline{M}_{\overline{q}},\overline{m}-N}+\overline{\epsilon}_{\overline{q}} & \left(N+1\leq\overline{m}\leq N+\overline{\mathcal{M}}_{\overline{q}}\right)\\
2\pi it_{\overline{M}_{\overline{q}-1},\overline{m}-N-\overline{\mathcal{M}}_{\overline{q}}}+\overline{\epsilon}_{\overline{q}-1} & \left(N+\overline{\mathcal{M}}_{\overline{q}}+1\leq\overline{m}\leq N+\overline{\mathcal{M}}_{\overline{q}-1}\right)\\
\,\,\,\,\,\,\,\,\vdots\\
2\pi it_{\overline{M}_{1},\overline{m}-N-\overline{\mathcal{M}}_{2}}+\overline{\epsilon}_{1} & \left(N+\overline{\mathcal{M}}_{2}+1\leq\overline{m}\leq N+\overline{\mathcal{M}}_{1}\right)
\end{cases},\label{eq:muRep}
\end{equation}
and
\begin{equation}
\nu_{m}=\begin{cases}
\beta_{m} & \left(1\leq m\leq N\right)\\
-2\pi it_{M_{q},m-N}+\epsilon_{q} & \left(N+1\leq m\leq N+\mathcal{M}_{q}\right)\\
-2\pi it_{M_{q-1},m-N-\mathcal{M}_{q}}+\epsilon_{q-1} & \left(N+\mathcal{M}_{q}+1\leq m\leq N+\mathcal{M}_{q-1}\right)\\
\,\,\,\,\,\,\,\,\vdots\\
-2\pi it_{M_{1},m-N-\mathcal{M}_{2}}+\epsilon_{1} & \left(N+\mathcal{M}_{2}+1\leq m\leq N+\mathcal{M}_{1}\right)
\end{cases}.\label{eq:nuRep}
\end{equation}

To proceed the computation, we divide this into a phase factor depending on $k$, a phase factor $\Omega$ independent of $k$, $Z_{0}$ which is related to the $N$-independent factor $C_{k,\boldsymbol{M}}^{\boldsymbol{s}}$ and $\mathcal{O}$ which is related to the density matrix:
\begin{equation}
\widetilde{Z}_{k,M}^{\left(+1,-1\right)}\left(N;\frac{\alpha}{\hbar},\frac{\beta}{\hbar}\right)=e^{i\Theta_{k,\bm{M}}^{\left(-\overline{q},+1,-q\right)}}\Omega Z_{0}\mathcal{O}.
\end{equation}
The explicit definitions are
\begin{align}
\Omega & =i^{-\frac{1}{2}\left\{ \left(N+\overline{\mathcal{M}}_{1}\right)^{2}-N^{2}\right\} +\frac{1}{2}\left\{ \left(N+\mathcal{M}_{1}\right)^{2}-N^{2}\right\} +\frac{\overline{\mathcal{M}}_{1}}{2}-\frac{\mathcal{M}_{1}}{2}},\nonumber \\
Z_{0} & =k^{-\frac{\overline{\mathcal{M}}_{1}}{2}-\frac{\mathcal{M}_{1}}{2}}\left(\prod_{\overline{a}}^{\overline{q}}\lim_{\overline{\epsilon}_{\overline{a}}\rightarrow0}\right)\left(\prod_{a}^{q}\lim_{\epsilon_{a}\rightarrow0}\right)\prod_{\overline{a}}^{\overline{q}-1}\prod_{\overline{j}_{\overline{a}+1}}^{\overline{\mathcal{M}}_{\overline{a}+1}}\frac{1}{2\cosh\frac{\mu_{N+\overline{j}_{\overline{a}+1}}-i\pi\overline{M}_{\overline{a}}}{2}}\nonumber \\
 & \,\,\,\times\frac{\prod_{\overline{j}<\overline{j}'}^{\overline{\mathcal{M}}_{1}}2\sinh\frac{\mu_{N+\overline{j}}-\mu_{N+\overline{j}'}}{2k}\prod_{j<j'}^{\mathcal{M}_{1}}2\sinh\frac{\nu_{N+j}-\nu_{N+j'}}{2k}}{\prod_{\overline{j}}^{\overline{\mathcal{M}}_{1}}\prod_{j}^{\mathcal{M}_{1}}2\cosh\frac{\mu_{N+\overline{j}}-\nu_{N+j}}{2k}}\prod_{a}^{q-1}\prod_{j_{a+1}}^{\mathcal{M}_{a+1}}\frac{1}{2\cosh\frac{\nu_{N+j_{a+1}}+i\pi M_{a}}{2}},\nonumber \\
\mathcal{O} & =\prod_{\overline{a}}^{\overline{q}}\prod_{\overline{n}}^{N}\frac{1}{2\cosh\frac{\mu_{\overline{n}}-i\pi\overline{M}_{\overline{a}}}{2}}\nonumber \\
 & \,\,\,\times\frac{\prod_{\overline{n}}^{N}\prod_{\overline{m}=\overline{n}+1}^{N+\overline{\mathcal{M}}_{1}}2\sinh\frac{\mu_{\overline{n}}-\mu_{\overline{m}}}{2k}\prod_{n}^{N}\prod_{m=n+1}^{N+\mathcal{M}_{1}}2\sinh\frac{\nu_{n}-\nu_{m}}{2k}}{\prod_{\left(\overline{m},m\right)\notin\left(N+\overline{j},N+j\right)}2\cosh\frac{\mu_{\overline{m}}-\nu_{m}}{2k}}\prod_{a}^{q}\prod_{n}^{N}\frac{1}{2\cosh\frac{\nu_{n}+i\pi M_{a}}{2}},
\end{align}
where $\prod_{\left(\overline{m},m\right)\notin\left(N+\overline{j},N+j\right)}$ means that $\overline{m}$ and $m$ run the range where $\overline{m}$ and $m$ are not larger than $N$ simultaneously. In other words, they run the following three ranges:
\begin{equation}
\begin{cases}
1\leq\overline{m}\leq N,\quad1\leq m\leq N\\
1\leq\overline{m}\leq N,\quad N+1\leq m\leq N+\mathcal{M}_{1}\\
N+1\leq\overline{m}\leq N+\overline{\mathcal{M}}_{1},\quad1\leq m\leq N
\end{cases}.
\end{equation}
 $Z_{0}$ is independent of $\alpha_{n}$ and $\beta_{n}$, while $\mathcal{O}$ depends on them. We start with $Z_{0}$ and especially focus on the factors independent of $\mu_{\overline{n}}$ (in other words, the factors depending only on $\nu_{n}$). We also put $k^{-\frac{\mathcal{M}_{1}}{2}}$ and the limit operation $\epsilon_{a}\rightarrow0$, and we denote by $I$ it. By substituting \eqref{eq:nuRep} and using the following identities
\begin{align}
\prod_{j<j'}^{\mathcal{M}_{1}}f\left(\nu_{N+j},\nu_{N+j'}\right) & =\prod_{a}^{q}\prod_{j_{a}<j_{a}'}^{M_{a}}f\left(-2\pi it_{M_{a},j_{a}}+\epsilon_{a},-2\pi it_{M_{a},j_{a}'}+\epsilon_{a}\right)\nonumber \\
 & \,\,\,\times\prod_{a}^{q-1}\prod_{b=a}^{q}\prod_{j_{a}}^{M_{a}}\prod_{j_{b}}^{M_{b}}f\left(-2\pi it_{M_{b},j_{b}}+\epsilon_{b},-2\pi it_{M_{a},j_{a}}+\epsilon_{a}\right),\nonumber \\
\prod_{j_{a}}^{\mathcal{M}_{a}}f\left(\nu_{N+j_{a}}\right) & =\prod_{b=a}^{q}\prod_{j_{b}}^{M_{b}}f\left(-2\pi it_{M_{b},j_{b}}+\epsilon_{b}\right),
\end{align}
we obtain
\begin{align}
I & =k^{-\frac{\mathcal{M}_{1}}{2}}\left(\prod_{a}^{q}\lim_{\epsilon_{a}\rightarrow0}\right)\left(\prod_{j<j'}^{\mathcal{M}_{1}}2\sinh\frac{\nu_{N+j}-\nu_{N+j'}}{2k}\right)\left(\prod_{a}^{q-1}\prod_{m_{a+1}=N+1}^{N+\mathcal{M}_{a+1}}\frac{1}{2\cosh\frac{\nu_{m_{a+1}}+i\pi M_{a}}{2}}\right)\nonumber \\
 & =\left(\prod_{a}^{q}i^{-\frac{1}{2}M_{a}\left(M_{a}-1\right)}Z_{k,M_{a}}^{\left(\mathrm{CS}\right)}\right)\left(\prod_{a}^{q}\lim_{\epsilon_{a}\rightarrow0}\right)\left(\prod_{a}^{q-1}\prod_{b=a+1}^{q}\prod_{j_{b}}^{M_{b}}\frac{\prod_{j_{a}}^{M_{a}}2\sinh\frac{-2\pi i\left(t_{M_{b},j_{b}}-t_{M_{a},j_{a}}\right)+\epsilon_{b}-\epsilon_{a}}{2k}}{2\cosh\frac{-2\pi it_{M_{b},j_{b}}+i\pi M_{a}+\epsilon_{b}}{2}}\right).
\end{align}
We now take the limit $\epsilon_{a}\rightarrow0$ because this operation is trivial (that is, even the fake divergences do not appear) for the other factors. The order of taking the limit is $a=1,2,\ldots,q$. The limit for $a=1$ is trivial. On the other hand, we should use the identity \eqref{eq:ZintForm} to take the limit for $a=2,3,\ldots,q$. We then get
\begin{equation}
I=i^{-\frac{1}{2}\mathcal{M}_{1}\left(\mathcal{M}_{1}-1\right)}\left(\prod_{a}^{q}Z_{k,M_{a}}^{\left(\mathrm{CS}\right)}\right)\left(\prod_{a}^{q-1}\prod_{b=a+1}^{q}Z_{k,M_{a},M_{b}}^{\left(\Iside\right)}\right).
\end{equation}
Next, we calculate the factors depending only on $\mu_{\overline{n}}$ with $k^{-\frac{\overline{\mathcal{M}}_{1}}{2}}$ and $\lim_{\overline{\epsilon}_{\overline{a}}\rightarrow0}$. The process is the same with the above, and the result is
\begin{align}
 & k^{-\frac{\overline{\mathcal{M}}_{1}}{2}}\left(\prod_{\overline{a}}^{\overline{q}}\lim_{\overline{\epsilon}_{\overline{a}}\rightarrow0}\right)\left(\prod_{\overline{j}<\overline{j}'}^{\overline{\mathcal{M}}_{1}}2\sinh\frac{\mu_{N+\overline{j}}-\mu_{N+\overline{j}'}}{2k}\right)\left(\prod_{\overline{a}}^{\overline{q}-1}\prod_{\overline{m}_{\overline{a}+1}=N+1}^{N+\overline{\mathcal{M}}_{\overline{a}+1}}\frac{1}{2\cosh\frac{\mu_{\overline{m}_{\overline{a}+1}}-i\pi\overline{M}_{\overline{a}}}{2}}\right)\nonumber \\
 & =i^{\frac{1}{2}\overline{\mathcal{M}}_{1}\left(\overline{\mathcal{M}}_{1}-1\right)}\left(\prod_{\overline{a}}^{\overline{q}}Z_{k,\overline{M}_{\overline{a}}}^{\left(\mathrm{CS}\right)}\right)\left(\prod_{\overline{a}}^{\overline{q}-1}\prod_{\overline{b}=\overline{a}+1}^{\overline{q}}Z_{k,\overline{M}_{\overline{a}},\overline{M}_{\overline{b}}}^{\left(\Iside\right)}\right).
\end{align}
Finally, we calculate the factors depending on both $\mu_{\overline{n}}$ and $\nu_{n}$:
\begin{equation}
\frac{1}{\prod_{\overline{j}}^{\overline{\mathcal{M}}_{1}}\prod_{j}^{\mathcal{M}_{1}}2\cosh\frac{\mu_{N+\overline{j}}-\nu_{N+j}}{2k}}=\prod_{\overline{a}}^{\overline{q}}\prod_{a}^{q}Z_{k,\overline{M}_{\overline{a}},M_{a}}^{\left(\Icross\right)}.
\end{equation}
In summary, $Z_{0}$ can be written as
\begin{align}
Z_{0} & =i^{\frac{1}{2}\overline{\mathcal{M}}_{1}\left(\overline{\mathcal{M}}_{1}-1\right)-\frac{1}{2}\mathcal{M}_{1}\left(\mathcal{M}_{1}-1\right)}\left(\prod_{\overline{a}}^{\overline{q}}Z_{k,\overline{M}_{\overline{a}}}^{\left(\mathrm{CS}\right)}\right)\left(\prod_{a}^{q}Z_{k,M_{a}}^{\left(\mathrm{CS}\right)}\right)\nonumber \\
 & \,\,\,\times\left(\prod_{\overline{a}}^{\overline{q}-1}\prod_{\overline{b}=\overline{a}+1}^{\overline{q}}Z_{k,\overline{M}_{\overline{a}},\overline{M}_{\overline{b}}}^{\left(\Iside\right)}\right)\left(\prod_{\overline{a}}^{\overline{q}}\prod_{a}^{q}Z_{k,\overline{M}_{\overline{a}},M_{a}}^{\left(\Icross\right)}\right)\left(\prod_{a}^{q-1}\prod_{b=a+1}^{q}Z_{k,M_{a},M_{b}}^{\left(\Iside\right)}\right).
\end{align}
Next, we combine $\Omega$ and the phase appeared in the above expression:
\begin{equation}
i^{\frac{1}{2}\overline{\mathcal{M}}_{1}\left(\overline{\mathcal{M}}_{1}-1\right)-\frac{1}{2}\mathcal{M}_{1}\left(\mathcal{M}_{1}-1\right)}\Omega=i^{N\left(\mathcal{M}_{1}-\overline{\mathcal{M}}_{1}\right)}.
\end{equation}
Finally, we combine this phase and $\mathcal{O}$. By substituting $\eqref{eq:muRep}$ and \eqref{eq:nuRep} and using the identities
\begin{align}
\prod_{\overline{m}}^{N+\overline{\mathcal{M}}_{1}}f\left(\mu_{\overline{m}}\right) & =\left(\prod_{\overline{n}}^{N}f\left(\alpha_{\overline{n}}\right)\right)\left(\prod_{\overline{a}}^{\overline{q}}\prod_{\overline{j}_{\overline{a}}}^{\overline{M}_{\overline{a}}}f\left(2\pi it_{\overline{M}_{\overline{a}},\overline{j}_{\overline{a}}}\right)\right),\nonumber \\
\prod_{m}^{N+\mathcal{M}_{1}}f\left(\nu_{m}\right) & =\left(\prod_{n}^{N}f\left(\beta_{n}\right)\right)\left(\prod_{a}^{q}\prod_{j_{a}}^{M_{a}}f\left(-2\pi it_{M_{a},j_{a}}\right)\right),
\end{align}
we obtain
\begin{align}
i^{N\left(\mathcal{M}_{1}-\overline{\mathcal{M}}_{1}\right)}\mathcal{O} & =\prod_{\overline{n}}^{N}\left\{ i^{\mathcal{M}_{1}-\overline{\mathcal{M}}_{1}}\left(\prod_{\overline{a}}^{\overline{q}}\frac{\prod_{\overline{j}_{\overline{a}}}^{\overline{M}_{\overline{a}}}2\sinh\frac{\alpha_{\overline{n}}-2\pi it_{\overline{M}_{\overline{a}},\overline{j}_{\overline{a}}}}{2k}}{2\cosh\frac{\alpha_{\overline{n}}-i\pi\overline{M}_{\overline{a}}}{2}}\right)\left(\prod_{a}^{q}\frac{1}{\prod_{j_{a}}^{M_{a}}2\cosh\frac{\alpha_{\overline{n}}-2\pi it_{M_{a},j_{a}}}{2k}}\right)\right\} \nonumber \\
 & \,\,\,\times\frac{\prod_{\overline{n}<\overline{n}'}^{N}2\sinh\frac{\alpha_{\overline{n}}-\alpha_{\overline{n}'}}{2k}\prod_{n<n'}^{N}2\sinh\frac{\beta_{n}-\beta_{n'}}{2k}}{\prod_{\overline{n}}^{N}\prod_{n}^{N}2\cosh\frac{\alpha_{\overline{n}}-\beta_{n}}{2k}}\nonumber \\
 & \,\,\,\times\prod_{n}^{N}\left\{ \left(\prod_{a}^{q}\frac{\prod_{j_{a}}^{M_{a}}2\sinh\frac{\beta_{n}-2\pi it_{M_{a},j_{a}}}{2k}}{2\cosh\frac{\beta_{n}+i\pi M_{a}}{2}}\right)\left(\prod_{\overline{a}}^{\overline{q}}\frac{1}{\prod_{\overline{j}_{\overline{a}}}^{\overline{M}_{\overline{a}}}2\cosh\frac{\beta_{n}-2\pi it_{\overline{M}_{\overline{a}},\overline{j}_{\overline{a}}}}{2}}\right)\right\} .\label{eq:Ogen}
\end{align}
Moreover, we use the determinant formula \eqref{eq:CauchyDet} and put the rest part into the determinant:
\begin{equation}
i^{N\left(\overline{\mathcal{M}}_{1}+\mathcal{M}_{1}\right)}\mathcal{O}=\hbar^{N}\det\left(\left[\bra{\alpha_{\overline{n}}}\widehat{\rho}_{\bm{M}}^{\left(-\overline{q},+1,-q\right)}\ket{\beta_{n}}\right]_{\overline{n},n}\right).
\end{equation}
Therefore, we finally obtain \eqref{eq:ZpGenRes}.

\subsection{Hanany-Witten transition\label{subsec:HWtrans}}

It is known that when an NS5-brane and a $\left(1,k\right)$5-brane pass through each other by moving along the $x^{6}$ direction, D3-branes stretched between them are created. This phenomenon is called Hanany-Witten transition \cite{Hanany:1996ie,Kitao:1998mf}. The duality of the worldvolume theories of brane configurations related by the Hanany-Witten transition is known as Seiberg-like duality \cite{Giveon:2008zn}. It is challenging to confirm this duality in the level of the partition functions. In this section, we show that the deformed factors \eqref{eq:ZpGenRes} of the dual theories are equal up to the phase. The equality between the deformed factors immediately leads to the equality between the whole matrix models, or equivalently, the partition functions.

In general, Hanany-Witten transition claims that when a $\left(p,q\right)$5-brane and a $\left(p',q'\right)$5-brane with $pq'-p'q\neq0$ pass through each other, $\left|pq'-p'q\right|$ D3-branes are created. In addition, this move reverses the orientation of D3-branes, and thus they become anti-D3-branes and annihilate with other D3-branes. In our case, when we start with a brane configuration
\begin{equation}
\left\langle N_{1}\circ N_{2}\bullet N_{3}\right\rangle ,
\end{equation}
and move the $\left(1,k\right)$5-brane across the NS5-brane, $k$ D3-branes are created, and thus the resulting brane configuration is
\begin{equation}
\left\langle N_{1}\bullet N_{1}+N_{3}-N_{2}+k\circ N_{3}\right\rangle .
\end{equation}
Therefore, in our setup, the two deformed factors associated with
\begin{align}
 & \left\langle 0\circ\overline{\mathcal{M}}_{\overline{q}}\circ\ldots\circ\overline{\mathcal{M}}_{2}\circ\overline{\mathcal{M}}_{1}\bullet\mathcal{M}_{1}\circ\ldots\circ\mathcal{M}_{q}\circ0\right\rangle _{N},\nonumber \\
 & \left\langle 0\circ\overline{\mathcal{M}}_{\overline{q}}\circ\ldots\circ\overline{\mathcal{M}}_{2}\bullet k-\overline{M}_{1}+\mathcal{M}_{1}\circ\mathcal{M}_{1}\circ\ldots\circ\mathcal{M}_{q}\circ0\right\rangle _{N},\label{eq:HWconfig}
\end{align}
should be equal. We introduce $\boldsymbol{M}^{\mathrm{origin}}$ and $\boldsymbol{M}^{\mathrm{HW}}$ denoting the relative ranks of the brane configurations in the first and the second line respectively. It was argued in \cite{Kapustin:2009kz} that the phase factors of the matrix models are attributed to using a framing which is different from the standard one. Therefore, we focus on the rest part
\begin{align}
 & e^{-i\Theta_{k,\boldsymbol{M}}^{\left(-\overline{q},+1,-q\right)}}\widetilde{Z}_{k,\boldsymbol{M}}^{\left(-\overline{q},+1,-q\right)}\left(N;\frac{\alpha}{\hbar},\frac{\beta}{\hbar}\right)\nonumber \\
 & =\left(\prod_{\overline{a}}^{\overline{q}}Z_{k,\overline{M}_{\overline{a}}}^{\left(\mathrm{CS}\right)}\right)\left(\prod_{a}^{q}Z_{k,M_{a}}^{\left(\mathrm{CS}\right)}\right)\nonumber \\
 & \,\,\,\times\left(\prod_{\overline{a}}^{\overline{q}-1}\prod_{\overline{b}=\overline{a}+1}^{\overline{q}}Z_{k,\overline{M}_{\overline{a}},\overline{M}_{\overline{b}}}^{\left(\Iside\right)}\right)\left(\prod_{\overline{a}}^{\overline{q}}\prod_{a}^{q}Z_{k,\overline{M}_{\overline{a}},M_{a}}^{\left(\Icross\right)}\right)\left(\prod_{a}^{q-1}\prod_{b=a+1}^{q}Z_{k,M_{a},M_{b}}^{\left(\Iside\right)}\right)\nonumber \\
 & \,\,\,\times\hbar^{N}\det\left(\left[\bra{\alpha_{\overline{n}}}\widehat{\rho}_{\boldsymbol{M}}^{\left(-\overline{q},+1,-q\right)}\ket{\beta_{n}}\right]_{\overline{n},n}^{N\times N}\right),\label{eq:PFabs}
\end{align}
and we prove that the phase removed deformed factors associated with \eqref{eq:HWconfig} are equal.

The important point is that the equality holds for the factors in each line of \eqref{eq:PFabs}, namely the pure Chern-Simons factors, the other $N$ independent factors depending on two relative ranks and the partition function of an ideal Fermi gas. To prove this, first we write the information of relative ranks in \eqref{eq:HWconfig} with the notation in the left side of \eqref{eq:McalToM}, namely difference notation
\begin{align}
\boldsymbol{M}^{\mathrm{origin}} & \rightarrow\left(\overline{M}_{\overline{q}},\ldots,\overline{M}_{2},\overline{M}_{1}|M_{1},M_{2},\ldots,M_{q}\right),\nonumber \\
\boldsymbol{M}^{\mathrm{HW}} & \rightarrow\left(\overline{M}_{\overline{q}},\ldots,\overline{M}_{2}|k-\overline{M}_{1},M_{1},M_{2},\ldots,M_{q}\right).\label{eq:HWlabel1}
\end{align}
This notation make it clear that the Hanany-Witten transition changes $\overline{M}_{1}$ in the left side to $k-\overline{M}_{1}$ in the right side. We prove the duality by focusing on this fact.

First, for the pure Chern-Simons factors, the Hanany-Witten transition changes $Z_{k,\overline{M}_{1}}^{\left(\mathrm{CS}\right)}$ to $Z_{k,k-\overline{M}_{1}}^{\left(\mathrm{CS}\right)}$. These values are equal as proved in appendix \ref{subsec:SLdual}. Second, for $Z_{k,M_{a},M_{b}}^{\left(\Iside\right)}$ and $Z_{k,\overline{M}_{\overline{a}},M_{a}}^{\left(\Icross\right)}$, the Hanany-Witten transition changes $Z_{k,\overline{M}_{1},\overline{M}_{\overline{a}}}^{\left(\Iside\right)}$ to $Z_{k,\overline{M}_{\overline{a}},k-\overline{M}_{1}}^{\left(\Icross\right)}$ and $Z_{k,\overline{M}_{1},M_{a}}^{\left(\Icross\right)}$ to $Z_{k,k-\overline{M}_{1},M_{a}}^{\left(\Iside\right)}$. These values are equal if the identity
\begin{equation}
Z_{k,\overline{M},M}^{\left(\Icross\right)}=Z_{k,k-\overline{M},M}^{\left(\Iside\right)},
\end{equation}
holds (consider also the second line of \eqref{eq:ZnormForm}). This identity actually holds, and the proof is also in appendix \ref{subsec:SLdual}. Third, for the density matrix, \eqref{eq:rhoGen} and \eqref{eq:SLDop} immediately lead
\begin{equation}
\widehat{\rho}_{\boldsymbol{M}^{\mathrm{origin}}}^{\left(-\overline{q},+1,-q\right)}=\widehat{\rho}_{\boldsymbol{M}^{\mathrm{HW}}}^{\left(-\left(\overline{q}-1\right),+1,-\left(q+1\right)\right)}.\label{eq:HWrho}
\end{equation}
The above three types of identities immediately lead to the equality of deformed factors of dual theories. We expect that identities we used above, or perhaps generalized ones, are also useful for other supersymmetric Chern-Simons theories listed at the end of section \ref{sec:Conclusion}.

We finally comment on the relation between the restriction \eqref{eq:ConfigCond} and the Hanany-Witten transition. \eqref{eq:HWlabel1} again makes it clear that the Hanany-Witten transition changes $\overline{M}_{1}+1\geq\overline{M}_{\overline{a}}$ to $k-\overline{M}_{1}+\overline{M}_{\overline{a}}\leq k+1$ and $M_{a}+\overline{M}_{1}\leq k+1$ to $k-\overline{M}_{1}+1\geq M_{a}$. These restrictions are equivalent as can be easily seen. Therefore, the restriction \eqref{eq:ConfigCond} is consistent with the Hanany-Witten transition.

\section{Quantum curves\label{sec:QC}}

The quantum curves at first naturally appeared as the inverses of the density matrices associated with the $\mathcal{N}=4$ Chern-Simons theories described by circular quiver diagrams without any rank deformations \cite{Marino:2011eh,Moriyama:2014gxa}. In this situation, namely $\boldsymbol{M}=\boldsymbol{0}$, the density matrices are the product of \eqref{eq:rhoTrivRes} following \eqref{eq:PFallRes} because we can cut the whole matrix models at all gauge nodes and decompose it completely as explained in section \ref{subsec:Strategy}. Therefore, the inverses of the density matrices are the products of $\left(\widehat{Q}^{\frac{1}{2}}+\widehat{Q}^{-\frac{1}{2}}\right)$ and $\left(\widehat{P}^{\frac{1}{2}}+\widehat{P}^{-\frac{1}{2}}\right)$, where
\begin{equation}
\widehat{Q}=e^{\widehat{q}},\quad\widehat{P}=e^{\widehat{p}}.
\end{equation}
Thus they are Laurent polynomials of $\widehat{Q}^{\frac{1}{2}}$ and $\widehat{P}^{\frac{1}{2}}$.

On the other hand, it is highly nontrivial that whether the inverses of rank deformed density matrices, which are denoted as
\begin{equation}
\widehat{H}_{\boldsymbol{M}}^{\boldsymbol{s}}=\left(\widehat{\rho}_{\boldsymbol{M}}^{\boldsymbol{s}}\right)^{-1},\label{eq:CurveDef}
\end{equation}
can be written as the specific form which can be termed the quantum curve or not. The progress for the ABJ theory gave a positive answer \cite{Kashaev:2015wia} (see also appendix A.3 in \cite{Kubo:2019ejc}) 
\begin{equation}
\widehat{H}_{M}^{\left(+1,-1\right)}=\widehat{\mathcal{Q}}_{-M}\widehat{P}^{\frac{1}{2}}+\widehat{\mathcal{Q}}_{M}\widehat{P}^{-\frac{1}{2}},\label{eq:QCabj}
\end{equation}
for the ABJ density matrix \eqref{eq:rhoABJ}, where
\begin{equation}
\widehat{\mathcal{Q}}_{M}=e^{\frac{1}{2}i\pi M}\widehat{Q}^{\frac{1}{2}}+e^{-\frac{1}{2}i\pi M}\widehat{Q}^{-\frac{1}{2}}.\label{eq:QpolDef}
\end{equation}
In the next section we show that the inverses of the density matrices \eqref{eq:rhoGen}, which we found in this paper, are also the quantum curves.

\subsection{Brane configurations and quantum curves\label{subsec:BCandQC}}

In this section, we relax the restrictions against the relative ranks \eqref{eq:ConfigCond}. We assume only that $M_{a}$ and $\overline{M}_{\overline{a}}$ are non-negative integers. For the right side and the left side of the labels of the relative ranks
\begin{equation}
\boldsymbol{M}=\left(\overline{\mathcal{M}}_{\overline{q}},\ldots,\overline{\mathcal{M}}_{1},\mathcal{M}_{1},\ldots,\mathcal{M}_{q}\right),
\end{equation}
we define the product of $\widehat{\mathcal{Q}}_{M}$ operator respectively:
\begin{equation}
\widehat{\mathcal{Q}}_{\bm{M}_{\mathrm{R}}}=\prod_{a}^{q}\widehat{\mathcal{Q}}_{M_{a}},\quad\widehat{\mathcal{Q}}_{\bm{M}_{\mathrm{L}}}=\prod_{\overline{a}}^{\overline{q}}\widehat{\mathcal{Q}}_{M_{\overline{a}}}.
\end{equation}
The inverses of the density matrices \eqref{eq:rhoGen} are
\begin{equation}
\widehat{H}_{\bm{M}}^{\left(-\overline{q},+1,-q\right)}=\widehat{\mathcal{Q}}_{-\bm{M}_{\mathrm{R}}}\widehat{P}^{\frac{1}{2}}\widehat{\mathcal{Q}}_{\bm{M}_{\mathrm{L}}}+\widehat{\mathcal{Q}}_{\bm{M}_{\mathrm{R}}}\widehat{P}^{-\frac{1}{2}}\widehat{\mathcal{Q}}_{-\bm{M}_{\mathrm{L}}}.\label{eq:QCgen}
\end{equation}
We prove this identity in appendix \ref{sec:QCproof}.

The above result means that the quantum curves associated with the brane configurations \eqref{eq:ConfiguGen} are \eqref{eq:QCgen}. This is natural in the following sense. As explained above, a quantum curve associated with a brane configuration without rank deformations can be obtained by replacing each NS5-brane and each $\left(1,k\right)$5-brane in the brane configuration to $\left(\widehat{P}^{\frac{1}{2}}+\widehat{P}^{-\frac{1}{2}}\right)$ and $\left(\widehat{Q}^{\frac{1}{2}}+\widehat{Q}^{-\frac{1}{2}}\right)$ respectively in reverse order (since quantum curves are the inverses of density matrices). Rank deformations break this factorization. However, for the ABJ theory case \eqref{eq:QCabj}, if we consider the term containing $\widehat{P}^{\frac{1}{2}}$ and the term containing $\widehat{P}^{-\frac{1}{2}}$ separately, the similar replacing rule can be applied. That is, the brane configuration of the ABJ theory is $\left\langle 0\bullet M\circ\right\rangle _{N}^{\mathrm{P}}$, and the ABJ quantum curve has the structure that $\widehat{\mathcal{Q}}_{\pm M}$ is located on the left side of $\widehat{P}^{\pm\frac{1}{2}}$. Our setup \eqref{eq:ConfiguGen} contains the larger number of $\left(1,k\right)$5-branes on both sides of the NS5-brane. \eqref{eq:QCgen} means that we can also apply a similar replacing rule to these brane configurations. That is, the corresponding quantum curves \eqref{eq:QCgen} have the structure that the same number of $\widehat{\mathcal{Q}}_{\pm M}$ as the $\left(1,k\right)$5-branes are located on both sides of $\widehat{P}^{\pm\frac{1}{2}}$.

The concept of order of operators in the above argument seems to be meaningless because two $\widehat{\mathcal{Q}}_{\pm M}$ are commutative and also we can exchange $\widehat{\mathcal{Q}}_{\pm M}$ and $\widehat{P}^{\pm\frac{1}{2}}$ using the identity
\begin{equation}
\widehat{P}^{\alpha}\widehat{\mathcal{Q}}_{M}=\widehat{\mathcal{Q}}_{M-2\alpha k}\widehat{P}^{\alpha},\label{eq:QmPexchange}
\end{equation}
where we used $\widehat{P}^{\alpha}\widehat{Q}^{\beta}=e^{-i\hbar\alpha\beta}\widehat{Q}^{\beta}\widehat{P}^{\alpha}$. However, conversely, we can consider that this implies dualities in the gauge theory side. In fact, the exchange of $\widehat{\mathcal{Q}}_{\pm M}$ and $\widehat{P}^{\pm\frac{1}{2}}$ is related to the Hanany-Witten transition as explained in the next paragraph. We also give a discussion about a ``Hanany-Witten move of two $\left(1,k\right)$5-branes'' in section \ref{sec:Conclusion}, which is related to the exchange of two $\widehat{\mathcal{Q}}_{\pm M}$.

We finally comment on the Hanany-Witten transition in terms of the quantum curves. The quantum curves associated with the dual theories are identical since the corresponding density matrices are identical \eqref{eq:HWrho}. Of course, this identity can be checked directly. Using \eqref{eq:QmPexchange}, we obtain
\begin{align}
\widehat{H}_{\boldsymbol{M}^{\mathrm{origin}}}^{\left(-\overline{q},+1,-q\right)} & =\widehat{\mathcal{Q}}_{-\bm{M}_{\mathrm{R}}^{\mathrm{origin}}}\widehat{P}^{\frac{1}{2}}\widehat{\mathcal{Q}}_{\overline{M}_{1}}\widehat{\mathcal{Q}}_{\bm{M}_{\mathrm{L}}^{\mathrm{HW}}}+\widehat{\mathcal{Q}}_{\bm{M}_{\mathrm{R}}^{\mathrm{origin}}}\widehat{P}^{-\frac{1}{2}}\widehat{\mathcal{Q}}_{-\overline{M}_{1}}\widehat{\mathcal{Q}}_{-\bm{M}_{\mathrm{L}}^{\mathrm{HW}}}\nonumber \\
 & =\widehat{\mathcal{Q}}_{-\bm{M}_{\mathrm{R}}^{\mathrm{origin}}}\widehat{\mathcal{Q}}_{-\left(k-\overline{M}_{1}\right)}\widehat{P}^{\frac{1}{2}}\widehat{\mathcal{Q}}_{\bm{M}_{\mathrm{L}}^{\mathrm{HW}}}+\widehat{\mathcal{Q}}_{\bm{M}_{\mathrm{R}}^{\mathrm{origin}}}\widehat{\mathcal{Q}}_{k-\overline{M}_{1}}\widehat{P}^{-\frac{1}{2}}\widehat{\mathcal{Q}}_{-\bm{M}_{\mathrm{L}}^{\mathrm{HW}}}\nonumber \\
 & =\widehat{H}_{\bm{M}^{\mathrm{HW}}}^{\left(-\left(\overline{q}-1\right),+1,-\left(q+1\right)\right)}.\label{eq:QCHW}
\end{align}
This identity reveals that the appearance of $k$ D3-branes in the Hanany-Witten move is translated into the noncommutativity between the position operator and the momentum operator.

\section{$\mathcal{N}=4$ Chern-Simons theories with four nodes\label{sec:D5Model}}

So far we studied the rank deformed gauge theories from three viewpoints, namely the viewpoints of brane configurations, density matrices and quantum curves. In this section we study the brane configurations consisting of two NS5-branes and two $\left(1,k\right)$5-branes using these results. Concretely, we focus on $\boldsymbol{s}=\left\{ +1,+1,-1,-1\right\} $ and $\boldsymbol{s}=\left\{ +1,-1,+1,-1\right\} $. The worldvolume theories of these brane configurations are the $\mathcal{N}=4$ Chern-Simons theories described by the circular quiver diagrams with the gauge groups
\begin{align}
 & \mathrm{U}\left(N_{1}\right)_{k}\times\mathrm{U}\left(N_{2}\right)_{0}\times\mathrm{U}\left(N_{3}\right)_{-k}\times\mathrm{U}\left(N_{4}\right)_{0},\nonumber \\
 & \mathrm{U}\left(N_{1}\right)_{k}\times\mathrm{U}\left(N_{2}\right)_{-k}\times\mathrm{U}\left(N_{3}\right)_{k}\times\mathrm{U}\left(N_{4}\right)_{-k}.
\end{align}
The first and the second theories are called $\left(2,2\right)$ theory and $\left(1,1,1,1\right)$ theory, respectively. Note that they are expected to share the same quantum curve since they are related by the Hanany-Witten transition.

The reason why we focus on these gauge theories is that motivated by the discovery of symmetries of these gauge theories by combining the Fermi gas formalism and numerical studies \cite{Moriyama:2014nca,Moriyama:2017gye,Moriyama:2017nbw}, a ``symmetry'' of the associated quantum curve has been studied \cite{Kubo:2018cqw}. They defined that quantum curves are symmetric if they are equal up to similarity transformations. This definition is natural because the values of the matrix models \eqref{eq:FGFform} are invariant under similarity transformations. They revealed that the associated quantum curve obeys the Weyl group symmetry of $\mathrm{SO}\left(10\right)$.

It was found in \cite{Kubo:2019ejc} that this discovery led to two important results. First, they found the explicit relations between the brane configurations associated with the $\left(2,2\right)$ theory and the $\left(1,1,1,1\right)$ theory and the associated quantum curve by using the symmetries in both sides, namely the Hanany-Witten transition and the $\mathrm{SL}\left(2,\mathbb{Z}\right)$ transformation in type IIB string theory and the Weyl group symmetry in the quantum curve, instead of using the Fermi gas approach. Second, in the quantum curve side, they found new symmetry which cannot be generated by the Hanany-Witten transition and the $\mathrm{SL}\left(2,\mathbb{Z}\right)$ transformation. In this section we review these two points and comment on them using our results.

\subsection{Quantum curves from Fermi gas approach\label{subsec:ProofOfConj}}

In this section at first we review the conjecture in \cite{Kubo:2019ejc} which predicts the explicit relation between the matrix model of the $\left(2,2\right)$ theory and a quantum curve, and then we partially prove this conjecture by using our results. We label the ranks of the $\left(2,2\right)$ theory as
\begin{equation}
\left\langle \widetilde{M}_{2}+\widetilde{M}_{3}\bullet\widetilde{M}_{1}+2\widetilde{M}_{3}\bullet2\widetilde{M}_{1}+\widetilde{M}_{2}+\widetilde{M}_{3}\circ\widetilde{M}_{1}\circ\right\rangle _{N}^{\mathrm{P}}.\label{eq:22Config}
\end{equation}
Note that $\widetilde{M}_{i}$ can be negative and half-integer. They conjectured that the corresponding quantum curve is
\begin{align}
\frac{\widehat{H}}{\alpha} & =\widehat{Q}\widehat{P}+\left(1+e_{2}e_{3}\right)\widehat{P}+e_{2}e_{3}\widehat{Q}^{-1}\widehat{P}\nonumber \\
 & \,\,\,+\left(1+e_{2}e_{3}^{-1}\right)\widehat{Q}+\frac{E}{\alpha}+e_{1}^{-1}e_{2}\left(e_{2}+e_{3}\right)\widehat{Q}^{-1}\nonumber \\
 & \,\,\,+e_{2}e_{3}^{-1}\widehat{Q}\widehat{P}^{-1}+e_{1}^{-1}e_{2}\left(e_{2}+e_{3}^{-1}\right)\widehat{P}^{-1}+e_{1}^{-2}e_{2}^{2}\widehat{Q}^{-1}\widehat{P}^{-1},\label{eq:22CurveConj}
\end{align}
where $e_{i}=e^{2\pi i\widetilde{M}_{i}}$.\footnote{In \cite{Kubo:2019ejc}, they indicated that it is necessary to fix one interval between two 5-branes for relating brane configurations and quantum curves. In this statement we implicitly put this ``reference'' on the left most D3-branes of the brane configuration in \eqref{eq:22Config}.} $\alpha$ and $E$ are not determined in the conjecture because these coefficients do not affect to the symmetry of the quantum curve. The direct derivation of the conjecture by using the Fermi gas approach is also important in this meaning because the Fermi gas approach determines also these coefficients. Note that the coefficients of $\widehat{Q}^{+1}$ and $\widehat{P}^{+1}$ are adjusted by the similarity transformations generated by $\widehat{Q}$ and $\widehat{P}$, which multiply each terms by constants according to $\widehat{P}^{\alpha}\widehat{Q}^{\beta}=e^{-i\hbar\alpha\beta}\widehat{Q}^{\beta}\widehat{P}^{\alpha}$.

So far we focused on the $\left(2,2\right)$ theory. Nevertheless, this conjecture also treats the $\left(1,1,1,1\right)$ theory because the Hanany-Witten transition relates the $\left(2,2\right)$ theory and the $\left(1,1,1,1\right)$ theory and the quantum curves are expected to obey the Hanany-Witten transition (this is the case for our setup \eqref{eq:QCHW}). Therefore, it is enough to study the $\left(2,2\right)$ theory.

In this paper we related the brane configurations to the quantum curves by using the Fermi gas approach. We now compare this relation to the conjecture \eqref{eq:22CurveConj} and show that these are consistent. Unfortunately, we cannot deal with all the rank deformations with the techniques we developed in this paper. The non-trivial brane configurations we can deal with are
\begin{align}
 & \left\langle M_{1}\bullet0\bullet M_{2}\circ0\circ\right\rangle _{N}^{\mathrm{P}},\nonumber \\
 & \left\langle 0\bullet0\bullet M_{1}+M_{2}\circ M_{2}\circ\right\rangle _{N}^{\mathrm{P}},\label{eq:22Config1}
\end{align}
where $M_{1}$ and $M_{2}$ are non-negative integers. Note that we can also deal with the $s=\pm1$ exchanged brane configurations by taking a complex conjugate. These brane configurations can be divided into two parts by cutting at two $0$ of the relative ranks. The quantum curves associated with the resulting parts are the special cases of \eqref{eq:QCgen}. The products of these quantum curves are the quantum curves associated with the whole brane configurations, which can be seen in \eqref{eq:PFallRes}. Note that we should be careful about the order of production. We should reverse the order since quantum curves are inverses of density matrices.

The quantum curve associated with the first line of \eqref{eq:22Config1} is
\begin{equation}
\widehat{H}_{M_{2}}^{\left(+1,-1\right)}\left(\widehat{H}_{M_{1}}^{\left(+1,-1\right)}\right)^{\dagger},
\end{equation}
where we took account of \eqref{eq:DensHermit}. This quantum curve has been calculated in \cite{Kubo:2019ejc} because the components are only the ABJ quantum curves, and they confirmed that the result is consistent with the conjecture. 

Next, we consider the second line of \eqref{eq:22Config1}. The corresponding quantum curve is
\begin{equation}
\widehat{H}_{\left(M_{1},M_{2}\right)}^{\left(+1,-2\right)}\widehat{H}^{\left(+1\right)}.
\end{equation}
We substitute \eqref{eq:QCgen}, take the similarity transformation $\widehat{P}^{-\frac{M_{2}}{2k}}\left[\cdot\right]\widehat{P}^{\frac{M_{2}}{2k}}$ (this operation changes $\widehat{Q}$ to $e^{i\pi M_{2}}\widehat{Q}$) and multiply it by $m_{1}^{\frac{1}{2}}m_{2}^{-\frac{1}{2}}$ where $m_{i}=e^{i\pi M_{i}}$:
\begin{align}
 & m_{1}^{\frac{1}{2}}m_{2}^{-\frac{1}{2}}\widehat{P}^{-\frac{M_{2}}{2k}}\widehat{H}_{\left(M_{1},M_{2}\right)}^{\left(+1,-2\right)}\widehat{H}^{\left(+1\right)}\widehat{P}^{\frac{M_{2}}{2k}}\nonumber \\
 & =\left\{ \left(\widehat{Q}^{\frac{1}{2}}+\widehat{Q}^{-\frac{1}{2}}\right)\left(\widehat{Q}^{\frac{1}{2}}+m_{1}m_{2}^{-1}\widehat{Q}^{-\frac{1}{2}}\right)\widehat{P}^{\frac{1}{2}}+\left(m_{1}m_{2}\widehat{Q}^{\frac{1}{2}}+\widehat{Q}^{-\frac{1}{2}}\right)\left(\widehat{Q}^{\frac{1}{2}}+m_{2}^{-2}\widehat{Q}^{-\frac{1}{2}}\right)\widehat{P}^{-\frac{1}{2}}\right\} \nonumber \\
 & \,\,\,\,\,\,\times\left(\widehat{P}^{\frac{1}{2}}+\widehat{P}^{-\frac{1}{2}}\right).
\end{align}
Now we return to the conjecture \eqref{eq:22CurveConj}. The brane configuration we focus on now is
\begin{equation}
\left(\widetilde{M}_{1},\widetilde{M}_{2},\widetilde{M}_{3}\right)=\left(\frac{1}{2}M_{1}+\frac{1}{2}M_{2},\frac{1}{2}M_{1},-\frac{1}{2}M_{2}\right),
\end{equation}
(and $N=-\frac{1}{2}M_{1}+\frac{1}{2}M_{2}$) in the notation of \eqref{eq:22Config}, so that the conjectured quantum curve is
\begin{align}
\frac{\widehat{H}}{\alpha} & =\widehat{Q}\widehat{P}+\left(1+m_{1}m_{2}^{-1}\right)\widehat{P}+m_{1}m_{2}^{-1}\widehat{Q}^{-1}\widehat{P}\nonumber \\
 & \,\,\,+\left(1+m_{1}m_{2}\right)\widehat{Q}+\frac{E}{\alpha}+m_{2}^{-1}\left(m_{1}+m_{2}^{-1}\right)\widehat{Q}^{-1}\nonumber \\
 & \,\,\,+m_{1}m_{2}\widehat{Q}\widehat{P}^{-1}+m_{2}^{-1}\left(m_{1}+m_{2}\right)\widehat{P}^{-1}+m_{2}^{-2}\widehat{Q}^{-1}\widehat{P}^{-1}.
\end{align}
These two quantum curves are exactly the same. Moreover, we find that $\alpha=m_{1}^{-\frac{1}{2}}m_{2}^{\frac{1}{2}}$ and $E=2m_{1}^{-\frac{1}{2}}m_{2}^{\frac{1}{2}}+2m_{1}^{\frac{1}{2}}m_{2}^{-\frac{1}{2}}$.

\subsection{Non-trivial symmetry\label{subsec:AccidentalSym}}

In \cite{Kubo:2019ejc}, they indicated that the two quantum curves associated with
\begin{align}
 & \left\langle N_{1}\bullet N_{2}\circ N_{3}\bullet N_{4}\circ\right\rangle ^{\mathrm{P}},\quad\left\langle N_{1}\bullet N_{3}\circ N_{2}\bullet N_{4}\circ\right\rangle ^{\mathrm{P}},\label{eq:ExpSymConfig}
\end{align}
are identical up to the similarity transformations based on the conjecture \eqref{eq:22CurveConj} (see (3.17) in \cite{Kubo:2019ejc}). This symmetry would be interesting because this symmetry cannot be generated by the Hanany-Witten transition and the $\mathrm{SL}\left(2,\mathbb{Z}\right)$ transformation. In this paper, we call this type of symmetry non-trivial symmetry. It would be natural to ask when the non-trivial symmetry arises. In this section, we try to answer this question by considering the non-trivial symmetry in terms of the density matrices. We focus on the special cases
\begin{align}
 & \left\langle 0\bullet\overline{M}\circ0\bullet M\circ\right\rangle _{N}^{\mathrm{P}},\quad\widehat{H}_{M}^{\left(+1,-1\right)}\widehat{H}_{\overline{M}}^{\left(+1,-1\right)},\nonumber \\
 & \left\langle 0\bullet0\circ\overline{M}\bullet M\circ\right\rangle _{N}^{\mathrm{P}},\quad\widehat{H}_{\overline{M},M}^{\left(-1,+1,-1\right)}\widehat{H}^{\left(+1\right)},\label{eq:ExpSymConfigEx}
\end{align}
to use our results. We wrote the corresponding quantum curves on the right side. The above explanation did not clarify the meaning of the non-trivial symmetry. Therefore, we explain this point at first. In \cite{Kubo:2018cqw}, they found that the two quantum curves are equal up to the similarity transformation:
\begin{equation}
\widehat{G}^{-1}\widehat{H}_{M}^{\left(+1,-1\right)}\widehat{H}_{\overline{M}}^{\left(+1,-1\right)}\widehat{G}=\widehat{H}_{\overline{M},M}^{\left(-1,+1,-1\right)}\widehat{H}^{\left(+1\right)},\label{eq:simQC1}
\end{equation}
where $\widehat{G}$ satisfies
\begin{equation}
\widehat{G}^{-1}\widehat{P}\widehat{G}=\widehat{\mathcal{Q}}_{-M-k}^{-1}\widehat{P}\widehat{\mathcal{Q}}_{M+k}.\label{eq:Sim1}
\end{equation}
This similarity transformation is non-trivial because we can only consider the similarity transformations which transform a polynomial to another polynomial, while this similarity transformation seems not to satisfy this restriction at first sight. It is for this reason we call the symmetry generated by $\widehat{G}$ non-trivial symmetry. In what follows we study the non-trivial symmetry using our results.

We first realize that $\widehat{G}$ is equal to $\widehat{G}_{M}$ defined in \eqref{eq:SimGen}. To understand the reason $\widehat{G}_{M}$ appears, we consider the inverse of \eqref{eq:simQC1} and then get the identity between the density matrices:
\begin{equation}
\widehat{G}_{\overline{M}}^{-1}\widehat{\rho}_{\overline{M}}^{\left(+1,-1\right)}\widehat{\rho}_{M}^{\left(+1,-1\right)}\widehat{G}_{\overline{M}}=\widehat{\rho}^{\left(+1\right)}\widehat{\rho}_{\overline{M},M}^{\left(-1,+1,-1\right)}.
\end{equation}
This identity is trivial because of \eqref{eq:rhoGen}, namely
\begin{align}
\widehat{\rho}_{\overline{M}}^{\left(+1,-1\right)}\widehat{\rho}_{M}^{\left(+1,-1\right)} & =\left(i^{\overline{M}}\frac{1}{\prod_{\overline{j}}^{\overline{M}}2\cosh\frac{\widehat{q}-2\pi it_{\overline{M},\overline{j}}}{2k}}\frac{1}{2\cosh\frac{\widehat{p}}{2}}\frac{\prod_{\overline{j}}^{\overline{M}}2\sinh\frac{\widehat{q}-2\pi it_{\overline{M},\overline{j}}}{2k}}{2\cosh\frac{\widehat{q}+i\pi\overline{M}}{2}}\right)\nonumber \\
 & \,\,\,\times\left(i^{M}\frac{1}{\prod_{j}^{M}2\cosh\frac{\widehat{q}-2\pi it_{M,j}}{2k}}\frac{1}{2\cosh\frac{\widehat{p}}{2}}\frac{\prod_{j}^{M}2\sinh\frac{\widehat{q}-2\pi it_{M,j}}{2k}}{2\cosh\frac{\widehat{q}+i\pi M}{2}}\right),\nonumber \\
\widehat{\rho}^{\left(+1\right)}\widehat{\rho}_{\overline{M},M}^{\left(-1,+1,-1\right)} & =\left(\frac{1}{2\cosh\frac{\widehat{p}}{2}}\right)\left(i^{M-\overline{M}}\frac{\prod_{\overline{j}}^{\overline{M}}2\sinh\frac{\widehat{q}-2\pi it_{\overline{M},\overline{j}}}{2k}}{2\cosh\frac{\widehat{q}-i\pi\overline{M}}{2}}\frac{1}{\prod_{j}^{M}2\cosh\frac{\widehat{q}-2\pi it_{M,j}}{2k}}\right.\nonumber \\
 & \,\,\,\left.\times\frac{1}{2\cosh\frac{\widehat{p}}{2}}\frac{\prod_{j}^{M}2\sinh\frac{\widehat{q}-2\pi it_{M,j}}{2k}}{2\cosh\frac{\widehat{q}+i\pi M}{2}}\frac{1}{\prod_{\overline{j}}^{\overline{M}}2\cosh\frac{\widehat{q}-2\pi it_{\overline{M},\overline{j}}}{2k}}\right).
\end{align}
That is, $\widehat{\rho}_{M}^{\left(+1,-1\right)}$ has the structure that $\left(2\cosh\frac{\widehat{p}}{2}\right)^{-1}$ lies at the center and $\left(\prod_{j}^{M}2\cosh\frac{\widehat{q}-2\pi it_{M,j}}{2k}\right)^{-1}$ and $\left(2\cosh\frac{\widehat{q}+i\pi M}{2}\right)^{-1}\times\left(\prod_{j}^{M}2\sinh\frac{\widehat{q}-2\pi it_{M,j}}{2k}\right)$ lie at the side, and $\widehat{G}_{\overline{M}}$ moves these of $\widehat{\rho}_{M}^{\left(+1,-1\right)}$ into $\widehat{\rho}_{\overline{M}}^{\left(+1,-1\right)}$. As a result, the length of one $\boldsymbol{s}=\left(+1,-1\right)$ decreases, namely $\widehat{\rho}_{\overline{M}}^{\left(+1,-1\right)}\rightarrow\widehat{\rho}^{\left(+1\right)}$, and the length of another $\boldsymbol{s}=\left(+1,-1\right)$ increases, namely $\widehat{\rho}_{M}^{\left(+1,-1\right)}\rightarrow\widehat{\rho}_{\overline{M},M}^{\left(-1,+1,-1\right)}$.

Based on the above argument, we realize that the non-trivial symmetry exists when the quantum curves are product of two $\widehat{\rho}_{\bm{M}}^{\left(-\overline{q},+1,-q\right)}$ since $\widehat{\rho}_{\bm{M}}^{\left(-\overline{q},+1,-q\right)}$ also has the structure that $\left(2\cosh\frac{\widehat{p}}{2}\right)^{-1}$ lies at the center and some $\left(\prod_{j}^{M}2\cosh\frac{\widehat{q}-2\pi it_{M,j}}{2k}\right)^{-1}$ and some $\left(2\cosh\frac{\widehat{q}+i\pi M}{2}\right)^{-1}\times\left(\prod_{j}^{M}2\sinh\frac{\widehat{q}-2\pi it_{M,j}}{2k}\right)$ lie at the side.

The above argument focuses only on the rank deformations which are the combination of two $\widehat{\rho}_{\bm{M}}^{\left(-\overline{q},+1,-q\right)}$. However, for example, the non-trivial symmetry exists for general rank deformations \eqref{eq:ExpSymConfig} in the $\left(1,1,1,1\right)$ theory. Therefore, we expect that our argument in this section can be applied also for general rank deformations. Thus we expect that the presence or the absence of the non-trivial symmetry depends only on the number of the NS5-branes and the $\left(1,k\right)$5-branes, and when the number of the NS5-branes (or the $\left(1,k\right)$5-branes) is just two, the non-trivial symmetry presents. In fact, the quantum curve associated with $\boldsymbol{s}=\left\{ +1,-1,-1+1,-1,-1\right\} $ obeys the Weyl group symmetry of $E_{7}$ \cite{Kubo:2018cqw}.

Note that the positive $N$ independent factors in \eqref{eq:ZpGenRes} do not obey the non-trivial symmetry. For example, the factors associated with the first line of \eqref{eq:ExpSymConfigEx} consist only of the pure Chern-Simons factors $Z_{k,\overline{M}}^{\left(\mathrm{CS}\right)}Z_{k,M}^{\left(\mathrm{CS}\right)}$, while the factors associated with the second line consist of $Z_{k,\overline{M},M}^{\left(\Icross\right)}$ in addition to the same pure Chern-Simons factors. Therefore, the non-trivial symmetry appears only in the partition functions of the ideal Fermi gases. In this meaning, it is essential for the non-trivial symmetry that the form of quiver diagrams is not linear but circular.

\section{Conclusion and discussion\label{sec:Conclusion}}

In this paper we applied the Fermi gas formalism to the matrix models computing the partition functions of $\mathcal{N}=4$ circular quiver Chern-Simons theories with various rank deformations. Our method can be applied to a wide class of rank deformations since we provided the computation method for deformed factors, which are parts of matrix models. After the computation, we found that the deformed factors factorize to the pure Chern-Simons factors, the other $N$ independent factors depending on two relative ranks and the partition functions of ideal Fermi gases. The equality between dual theories related by the Hanany-Witten transition holds for each type of factor. We also found that the inverses of the density matrices can be rewritten as the quantum curves. Using these results we studied specific $\mathcal{N}=4$ Chern-Simons theories with the gauge group consisting of four unitary groups, and we especially focused on the non-trivial symmetry generated by \eqref{eq:SimGen}.

The quantum curves obey this non-trivial symmetry, while the $N$ independent factors do not obey it as explained in section \ref{subsec:AccidentalSym}. Interestingly, the same thing happens when we consider the exchange of two $M_{a}$ or two $\overline{M}_{a}$ in the left side notation of \eqref{eq:McalToM}, namely difference notation. This is because the quantum curves \eqref{eq:QCgen} of before and after the exchange are completely equal as can be easily seen, while $Z_{k,M,M'}^{\left(\Iside\right)}\neq Z_{k,M',M}^{\left(\Iside\right)}$ in general.\footnote{The parameters of rank deformations on the left-hand side or the right-hand side of the inequality may not satisfy the restriction in the second line of \eqref{eq:ConfigCond}. However, we can calculate both sides when the difference between the two parameters is one. The result is $Z_{k,M+1,M}^{\left(\Iside\right)}=2Z_{k,M,M+1}^{\left(\Iside\right)}$.} We might be able to relate this symmetry to a ``Hanany-Witten move of two $\left(1,k\right)$5-branes''. Though the argument developed in \cite{Hanany:1996ie} is not applied for the exchange of 5-branes of the same type, here we assume that the argument in section \ref{subsec:HWtrans} also holds in this case. Then, since no D3-branes are created, the exchange of two adjacent $\left(1,k\right)$5-branes in the brane configuration \eqref{eq:ConfiguGen} has the effect of exchanging two adjacent $M_{a}$ or two adjacent $\overline{M}_{a}$. In any case, it would be great to reveal the physical origin of the symmetries obeyed by the quantum curves (or equivalently the density matrices) but not obeyed by the $N$ independent factors.

We shall list some further directions in the following. First, we performed the computation of the deformed factors under the restriction \eqref{eq:ConfigCond}. The first line of this restriction, namely $M_{a}\leq k$ and $\overline{M}_{a}\leq k$ are just for using the argument in appendix \ref{sec:integral} from the technical point of view. It is interesting to uncover the behavior of the matrix models in the Fermi gas description beyond the restriction.

Second, as explained in the introduction, our main motivation to consider the Fermi gas formalism is the expectation that using the formalism we can calculate, for example, the membrane instantons and the exact values of the matrix models associated with the rank deformed $\mathcal{N}=4$ Chern-Simons theories as with the case of the ABJ(M) theory and the $\mathcal{N}=4$ Chern-Simons theories without rank deformations. We leave these problems for future work. We also expect that we can analyze physical values using the Fermi gas formalism. In particular, it is known in the ABJ(M) theory that on the one hand the vacuum expectation values of the Wilson loops are interpreted to be the insertion of corresponding operators to the partition function of the ABJ(M) Fermi gas \cite{Klemm:2012ii,Hatsuda:2013yua}, but on the other hand they are interpreted to be the deformations of the ABJ(M) density matrix \cite{Kiyoshige:2016lno} in harmony with the rank deformations. The latter interpretation is interesting because this interpretation deeply related to the open-closed duality of the topological string theory on local $\mathbb{P}^{1}\times\mathbb{P}^{1}$ \cite{Hatsuda:2016rmv,Kiyoshige:2016lno}. It is interesting whether a similar story exists in more general cases we studied.

Third, in this paper we focused on the specific brane configurations \eqref{eq:ConfiguGen}, and therefore we need to develop the computation technique to deal with more general brane configurations. Furthermore, there are various generalizations of the ABJM theory, and it is expected that the Fermi gas formalism for these theories with rank deformations provide, for example, identities between the partition functions of dual theories. Therefore, we hope to apply the Fermi formalism to, for example, mass deformed theories \cite{Hosomichi:2008jd} (the result for the ABJM theory is in \cite{Nosaka:2015iiw}), theories with gauge groups including Lie groups other than unitary groups (the result for the gauge group consisting of two groups is in \cite{Moriyama:2016xin}), theories including the fundamental matters \cite{Marino:2011eh,Grassi:2014vwa}, and the D-type quiver theories (the results for specific rank theories are in \cite{Assel:2015hsa,Moriyama:2015jsa}) and the E-type quiver theories in the context of the ADE classification \cite{Gulotta:2011vp}.

\section*{Acknowledgments}

We are grateful to Masazumi Honda, Seiji Terashima, Koji Umemoto, Shuichi Yokoyama and especially Sanefumi Moriyama for valuable discussions and comments. This work was supported by Grant-in-Aid for JSPS Fellows No.20J12263.

\appendix

\section{Formulas\label{sec:Formulas}}

This appendix provides various formulas we use in our computations.

\subsection{Determinant formula}

In this section we show that the 1-loop contributions for each 5-brane \eqref{eq:OneLoop} can be written as a determinant of a matrix all whose elements are inner products of vectors. Explicitly, the following identities hold for non-negative integer $N$ and $M$:

\begin{align}
 & \frac{\prod_{n<n'}^{N}2\sinh\frac{\alpha_{n}-\alpha_{n'}}{2k}\prod_{m<m'}^{N+M}2\sinh\frac{\beta_{m}-\beta_{m'}}{2k}}{\prod_{n}^{N}\prod_{m}^{N+M}2\cosh\frac{\alpha_{n}-\beta_{m}}{2k}}\nonumber \\
 & =\det\left(\begin{array}{c}
\left[\hbar\bra{\alpha_{n}}\frac{1}{2\cosh\frac{\widehat{p}-i\pi M}{2}}\ket{\beta_{m}}\right]_{n,m}^{N\times\left(N+M\right)}\\
\left[\frac{\hbar}{\sqrt{k}}\bbraket{2\pi it_{M,j}|\beta_{m}}\right]_{j,m}^{M\times\left(N+M\right)}
\end{array}\right),\nonumber \\
 & \frac{\prod_{m<m'}^{N+M}2\sinh\frac{\alpha_{m}-\alpha_{m'}}{2k}\prod_{n<n'}^{N}2\sinh\frac{\beta_{n}-\beta_{n'}}{2k}}{\prod_{m}^{N+M}\prod_{n}^{N}2\cosh\frac{\alpha_{m}-\beta_{n}}{2k}}\nonumber \\
 & =\det\left(\begin{array}{cc}
\left[\hbar\bra{\alpha_{m}}\frac{1}{2\cosh\frac{\widehat{p}+i\pi M}{2}}\ket{\beta_{n}}\right]_{m,n}^{\left(N+M\right)\times N} & \left[\frac{\hbar}{\sqrt{k}}\brakket{\alpha_{m}|-2\pi it_{M,j}}\right]_{m,j}^{\left(N+M\right)\times M}\end{array}\right),\label{eq:CauchyDet}
\end{align}
where $\left[f_{a,b}\right]_{a,b}^{A\times B}$ denotes an $A\times B$ matrix whose $\left(a,b\right)$ element is $f_{a,b}$, and $t_{M,j}$ is defined in \eqref{eq:tMdef}. We only derive the first identity in the rest of this section because the second identity can be easily derived from the first identity.

The starting point is the combination of the Cauchy determinant formula and the Vandermonde determinant formula \cite{Matsumoto:2013nya,Assel:2014awa}:
\begin{align}
\frac{\prod_{n<n'}^{N}2\sinh\frac{\alpha_{n}-\alpha_{n'}}{2k}\prod_{m<m'}^{N+M}2\sinh\frac{\beta_{m}-\beta_{m'}}{2k}}{\prod_{n}^{N}\prod_{m}^{N+M}2\cosh\frac{\alpha_{n}-\beta_{m}}{2k}} & =\left(-1\right)^{NM}\det\left(\begin{array}{c}
\left[\frac{e^{\frac{M}{2k}\left(\alpha_{n}-\beta_{m}\right)}}{2\cosh\frac{\alpha_{n}-\beta_{m}}{2k}}\right]_{n,m}^{N\times\left(N+M\right)}\\
\left[e^{\frac{1}{k}t_{M,j}\beta_{m}}\right]_{j,m}^{M\times\left(N+M\right)}
\end{array}\right).\label{eq:Det0}
\end{align}
We calculate the upper part of the matrix in the right-hand side. Using the Fourier transform
\begin{equation}
\frac{1}{2\cosh\pi x}=\frac{1}{2\pi}\int_{\mathbb{R}}dp\frac{e^{ipx}}{2\cosh\frac{p}{2}},
\end{equation}
we get
\begin{align}
\frac{e^{\frac{M}{2k}\left(\alpha-\beta\right)}}{2\cosh\frac{\alpha-\beta}{2k}} & =\frac{1}{2\pi}\int_{\mathbb{R}}dp\frac{e^{\frac{i}{2\pi k}\left(p-i\pi M\right)\left(\alpha-\beta\right)}}{2\cosh\frac{p}{2}}\nonumber \\
 & =\frac{1}{2\pi}\int_{\mathbb{R}}dp\frac{e^{\frac{i}{2\pi k}p\left(\alpha-\beta\right)}}{2\cosh\frac{p+i\pi M}{2}}+\sum_{j}^{\left\lfloor \frac{M+1}{2}\right\rfloor }\left(-1\right)^{j+1}e^{\frac{1}{k}t_{M,j}\left(\alpha-\beta\right)}.
\end{align}
At the second line we shifted the integration contour from $\mathbb{R}$ to $\mathbb{R}+i\pi M$. The terms in the sum in the second line are residues of $\left(2\cosh\frac{p}{2}\right)^{-1}$, which has poles at $p=\left(2j-1\right)\pi i$ with residues $\left(-1\right)^{j}i$ .\footnote{There are two subtleties when $M$ is odd, because the shifted contour passes over the pole at $j=\frac{M+1}{2}$. First, the integrand has the pole at $p=0$. This integrand becomes $2\cosh\frac{\alpha_{\overline{n}}-i\pi\overline{M}_{\overline{a}}}{2}$ or $2\cosh\frac{\beta_{n}+i\pi M_{a}}{2}$ in the denominator of \eqref{eq:Ogen}. The zeros at $\alpha_{\overline{n}}=0$ and $\beta_{n}=0$ cancel with zeros appeared in the numerator. In this meaning, we do not mind this problem. Second, the residue of $j=\frac{M+1}{2}$ might be the half. However, this term vanishes as explained just below. Therefore, we also do not mind this problem.} However, these terms do not contribute to the determinant in \eqref{eq:Det0} because these terms can be written as linear combinations of the lower part of the matrix. Finally, we reach \eqref{eq:CauchyDet} using \eqref{eq:Normalization}.

\subsection{$N$ independent factors\label{subsec:NormalizationConst}}

In this appendix we provide identities  related to the positive $N$-independent factors \eqref{eq:ZnormDef}.

First, the identities
\begin{align}
Z_{k,M,M}^{\left(\Iside\right)} & =\left(Z_{k,M}^{\left(\mathrm{CS}\right)}\right)^{2},\nonumber \\
Z_{k,\overline{M},M}^{\left(\Icross\right)} & =Z_{k,M,\overline{M}}^{\left(\Icross\right)},\label{eq:ZnormForm}
\end{align}
easily follow from their definitions. Second, the following identity holds for non-negative integer $M$ and $M'$ satisfying $M'\leq M+1$:
\begin{align}
\lim_{\epsilon\rightarrow0}\prod_{j'}^{M'}\left\{ \left(\prod_{j}^{M}2\sinh\frac{2\pi i\left(t_{M',j'}-t_{M,j}\right)+\epsilon}{2k}\right)\frac{1}{2\cosh\frac{2\pi it_{M',j'}+i\pi M+\epsilon}{2}}\right\}  & =i^{-MM'}Z_{k,M,M'}^{\left(\Iside\right)}.\label{eq:ZintForm}
\end{align}
In the rest of this section, we prove this identity

The left-hand side of \eqref{eq:ZintForm}, which is denoted by $I$, is equal to
\begin{equation}
I=i^{-MM'}\prod_{j'}^{M'}\left\{ \lim_{\epsilon_{j'}\rightarrow0}\left(\prod_{j}^{M}2\sinh\frac{2\pi i\left(t_{M',j'}-t_{M,j}\right)+\epsilon_{j'}}{2k}\right)\frac{1}{e^{\frac{\epsilon_{j'}}{2}}+e^{-\pi i\left(2t_{M',j'}+M\right)}e^{-\frac{\epsilon_{j'}}{2}}}\right\} .
\end{equation}
At this point, we consider separately the cases when $M+M'$ is even or odd. First, we consider the even case. In this case, zero appears in the denominator after taking the limit, because $2t_{M',j'}+M$ is odd. However, the restriction $M'\leq M+1$ ensures the appearance of zero also in the numerator. Therefore,
\begin{align}
I & =i^{-MM'}\prod_{j'}^{M'}\left\{ \left(\prod_{t_{M,j}\neq t_{M',j'}}^{M}2\sinh\frac{2\pi i\left(t_{M',j'}-t_{M,j}\right)}{2k}\right)\lim_{\epsilon_{j'}\rightarrow0}\frac{2\sinh\frac{\epsilon_{j'}}{2k}}{2\sinh\frac{\epsilon_{j'}}{2}}\right\} \nonumber \\
 & =i^{-MM'}Z_{k,M,M'}^{\left(\Iside\right)},
\end{align}
where $\prod_{t_{M,j}\neq t_{M',j'}}^{M}$ means that $j$ runs from $1$ to $M$ except for the case when $t_{M,j}=t_{M',j'}$. We would like to emphasize that $I$ diverges if $M+M'$ is even and $M'\geq M+2$. Second, we consider the odd case. In this case, we can take the limit trivially because $2t_{M',j'}+M$ is even, and we again obtain \eqref{eq:ZintForm}.

\subsection{Hanany-Witten transition\label{subsec:SLdual}}

In this appendix we provide identities related to the Hanany-Witten transition. We assume that non-negative integers $k$, $M$ and $\overline{M}$ satisfy $M\leq k$, $\overline{M}\leq k$ and $M+\overline{M}\leq k+1$. The identities related to the positive $N$-independent factors \eqref{eq:ZnormDef} are
\begin{align}
Z_{k,M}^{\left(\mathrm{CS}\right)} & =Z_{k,k-M}^{\left(\mathrm{CS}\right)},\nonumber \\
Z_{k,\overline{M},M}^{\left(\Icross\right)} & =Z_{k,k-\overline{M},M}^{\left(\Iside\right)},\label{eq:SLDnorm}
\end{align}
and related to the density matrices are
\begin{align}
 & i^{M}\frac{1}{\prod_{j}^{M}2\cosh\frac{\widehat{q}-2\pi it_{M,j}}{2k}}\frac{1}{2\cosh\frac{\widehat{p}}{2}}\frac{\prod_{j}^{M}2\sinh\frac{\widehat{q}-2\pi it_{M,j}}{2k}}{2\cosh\frac{\widehat{q}+i\pi M}{2}}\nonumber \\
 & =i^{-\left(k-M\right)}\frac{\prod_{j}^{k-M}2\sinh\frac{\widehat{q}-2\pi it_{k-M,j}}{2k}}{2\cosh\frac{\widehat{q}-i\pi\left(k-M\right)}{2}}\frac{1}{2\cosh\frac{\widehat{p}}{2}}\frac{1}{\prod_{j}^{k-M}2\cosh\frac{\widehat{q}-2\pi it_{k-M,j}}{2k}}.\label{eq:SLDop}
\end{align}
In the rest of this section, we prove these identities.\footnote{The identity in the first line of \eqref{eq:SLDnorm} was proved in \cite{Kapustin:2010mh}.}

We start with the well-known identity
\begin{equation}
z^{k}-1=\prod_{j}^{k}\left(z-e^{\frac{2\pi i}{k}j}\right).
\end{equation}
By substituting $z=e^{-\frac{1}{k}\left(q-i\pi\left(M+1\right)\right)}$, we get
\begin{equation}
i^{k}\left(-1\right)^{M}2\cosh\frac{q+i\pi M}{2}=\prod_{j}^{k}2\sinh\frac{q-2i\pi t_{M,j}}{2k}.
\end{equation}
We divide the production in the right-hand side into less than and not less than $M+1$. The latter part becomes
\begin{equation}
\prod_{j=M+1}^{k}2\sinh\frac{q-2i\pi t_{M,j}}{2k}=i^{k-M}\prod_{j=1}^{k-M}2\cosh\frac{q-2i\pi t_{k-M,j}}{2k}.
\end{equation}
Hence we obtain
\begin{equation}
i^{-M}2\cosh\frac{q+i\pi M}{2}=\left(\prod_{j}^{M}2\sinh\frac{q-2\pi it_{M,j}}{2k}\right)\left(\prod_{j}^{k-M}2\cosh\frac{q-2\pi it_{k-M,j}}{2k}\right).\label{eq:OpDual}
\end{equation}
The combination of this identity and the identity obtained by changing $M$ into $k-M$ in this identity leads \eqref{eq:SLDop}.

Furthermore, by changing $M$ into $k-\overline{M}$ in \eqref{eq:OpDual} we get
\begin{equation}
i^{k-\overline{M}}\frac{\prod_{\overline{j}}^{k-\overline{M}}2\sinh\frac{q-2\pi it_{k-\overline{M},\overline{j}}}{2k}}{2\cosh\frac{q+i\pi\left(k-\overline{M}\right)}{2}}=\prod_{\overline{j}}^{\overline{M}}\frac{1}{2\cosh\frac{q-2\pi it_{\overline{M},\overline{j}}}{2k}}.
\end{equation}
By substituting $q=2\pi it_{M,j}+\epsilon$ and multiplying $j=1,2,\ldots,M$ we get
\begin{equation}
i^{\left(k-\overline{M}\right)M}\prod_{j}^{M}\frac{\prod_{\overline{j}}^{k-\overline{M}}2\sinh\frac{2\pi it_{M,j}-2\pi it_{k-\overline{M},\overline{j}}+\epsilon}{2k}}{2\cosh\frac{2\pi it_{M,j}+i\pi\left(k-\overline{M}\right)+\epsilon}{2}}=\prod_{\overline{j}}^{\overline{M}}\prod_{j}^{M}\frac{1}{2\cosh\frac{2\pi it_{\overline{M},\overline{j}}-2\pi it_{M,j}-\epsilon}{2k}}.
\end{equation}
We take the limit $\epsilon\rightarrow0$ using \eqref{eq:ZintForm} for the left-hand side and the inequality $\left|t_{\overline{M},\overline{j}}-t_{M,j}\right|\leq\frac{k-1}{2}$ for the right-hand side, and we obtain the identity in the second line of \eqref{eq:SLDnorm}.

Finally, using \eqref{eq:ZnormForm} and \eqref{eq:SLDnorm} we get
\begin{equation}
\left(Z_{k,M}^{\left(\mathrm{CS}\right)}\right)^{2}=Z_{k,M,M}^{\left(\Iside\right)}=Z_{k,k-M,M}^{\left(\Icross\right)}=Z_{k,M,k-M}^{\left(\Icross\right)}=Z_{k,k-M,k-M}^{\left(\Iside\right)}=\left(Z_{k,k-M}^{\left(\mathrm{CS}\right)}\right)^{2}.
\end{equation}
Moreover, $Z_{k,M}^{\left(\mathrm{CS}\right)}$ is positive. Therefore, the identity in the first line of \eqref{eq:SLDnorm} holds.

\subsection{Similarity transformation}

In this section we prove the identity
\begin{align}
 & \widehat{G}_{M}^{-1}\widehat{P}\widehat{G}_{M}=\frac{1}{\widehat{\mathcal{Q}}_{-M-k}}\widehat{P}\widehat{\mathcal{Q}}_{M+k},\nonumber \\
 & \widehat{G}_{M}\widehat{P}\widehat{G}_{M}^{-1}=\widehat{\mathcal{Q}}_{-M-k}\widehat{P}\frac{1}{\widehat{\mathcal{Q}}_{M+k}},\label{eq:LargeSim}
\end{align}
where $\widehat{\mathcal{Q}}_{M}$ is defined in \eqref{eq:QpolDef}, and
\begin{equation}
\widehat{G}_{M}=\prod_{j}^{M}\frac{1}{2\cosh\frac{\widehat{q}-2\pi it_{M,j}}{2k}}.\label{eq:SimGen}
\end{equation}
The identity in the second line of \eqref{eq:LargeSim} is the Hermitian conjugate of the first line. In the rest of this section, we prove the first line.

The quantum dilogarithm $\Phi_{\mathsf{b}}\left(x\right)$ plays an important role. This function can be expressed as \cite{Faddeev:1993rs,Faddeev:1995nb,Faddeev:2000if}
\begin{equation}
\Phi_{\mathsf{b}}\left(x\right)=\frac{\left(e^{2\pi\mathsf{b}\left(x+\frac{i}{2}\left(\mathsf{b}+\mathsf{b}^{-1}\right)\right)};e^{2\pi i\mathsf{b}^{2}}\right)_{\infty}}{\left(e^{2\pi\mathsf{b}^{-1}\left(x-\frac{i}{2}\left(\mathsf{b}+\mathsf{b}^{-1}\right)\right)};e^{-2\pi i\mathsf{b}^{-2}}\right)_{\infty}},
\end{equation}
where
\begin{equation}
\left(a;q\right)_{n}=\prod_{k=0}^{n-1}\left(1-aq^{k}\right),
\end{equation}
is the q-Pochhammer symbol. The quantum dilogarithm satisfies
\begin{align}
\frac{\Phi_{\mathsf{b}}\left(x+\frac{i}{2}\mathsf{b}\right)}{\Phi_{\mathsf{b}}\left(x-\frac{i}{2}\mathsf{b}\right)} & =\frac{1}{1+e^{2\pi\mathsf{b}x}},\nonumber \\
\frac{\Phi_{\mathsf{b}}\left(x+\frac{i}{2}\mathsf{b}^{-1}\right)}{\Phi_{\mathsf{b}}\left(x-\frac{i}{2}\mathsf{b}^{-1}\right)} & =\frac{1}{1+e^{2\pi\mathsf{b}^{-1}x}}.\label{eq:DilogRatio}
\end{align}
We substitute $x=\frac{q-2\pi it_{M,j}}{2\pi\sqrt{k}}$ to the first line of \eqref{eq:DilogRatio} and multiply $j=1,2,\ldots,M$ , then we obtain
\begin{equation}
\frac{\Phi_{\sqrt{k}}\left(\frac{q}{2\pi\sqrt{k}}+\frac{iM}{2\sqrt{k}}\right)}{\Phi_{\sqrt{k}}\left(\frac{q}{2\pi\sqrt{k}}-\frac{iM}{2\sqrt{k}}\right)}=e^{-\frac{M}{2k}q}\prod_{j}^{M}\frac{1}{2\cosh\frac{q-2\pi it_{M,j}}{2k}}.
\end{equation}
Moreover, using the second line of \eqref{eq:DilogRatio} and $f\left(\widehat{q}\right)\widehat{P}=\widehat{P}f\left(\widehat{q}+2\pi ik\right)$, we obtain
\begin{align}
 & \frac{\Phi_{\sqrt{k}}\left(\frac{\widehat{q}}{2\pi\sqrt{k}}-\frac{iM}{2\sqrt{k}}\right)}{\Phi_{\sqrt{k}}\left(\frac{\widehat{q}}{2\pi\sqrt{k}}+\frac{iM}{2\sqrt{k}}\right)}\widehat{P}\frac{\Phi_{\sqrt{k}}\left(\frac{\widehat{q}}{2\pi\sqrt{k}}+\frac{iM}{2\sqrt{k}}\right)}{\Phi_{\sqrt{k}}\left(\frac{\widehat{q}}{2\pi\sqrt{k}}-\frac{iM}{2\sqrt{k}}\right)}\nonumber \\
 & =\frac{\Phi_{\sqrt{k}}\left(\frac{\widehat{q}}{2\pi\sqrt{k}}-\frac{iM}{2\sqrt{k}}\right)}{\Phi_{\sqrt{k}}\left(\frac{\widehat{q}}{2\pi\sqrt{k}}-\frac{iM}{2\sqrt{k}}-i\sqrt{k}\right)}\widehat{P}\frac{\Phi_{\sqrt{k}}\left(\frac{\widehat{q}}{2\pi\sqrt{k}}+\frac{iM}{2\sqrt{k}}\right)}{\Phi_{\sqrt{k}}\left(\frac{\widehat{q}}{2\pi\sqrt{k}}+\frac{iM}{2\sqrt{k}}+i\sqrt{k}\right)}\nonumber \\
 & =\frac{1}{1+e^{-i\pi\left(M+k\right)}\widehat{Q}}\widehat{P}\left(1+e^{i\pi\left(M+k\right)}\widehat{Q}\right).
\end{align}
The combination of these two identities leads \eqref{eq:LargeSim}.

\section{Complex delta functions\label{sec:integral}}

In this appendix we briefly comment on the integral
\begin{equation}
\int_{\mathbb{R}}dxf\left(x\right)\braket{x|a+ib},\label{eq:CompDelta}
\end{equation}
where $a$ and $b$ are real numbers. If $f\left(x\right)=o\left(e^{-\left|x\right|}\right)$ and $f\left(x\right)$ has neither poles nor branch points in $0<\mathrm{Im}\left(x\right)<b$ or $b<\mathrm{Im}\left(x\right)<0$ when $b>0$ or $b<0$ respectively, $\braket{x|a+ib}$ is considered to be the usual delta function. Concretely, we can perform the integration by shifting the integration contour from $\mathbb{R}$ to $\mathbb{R}+ib$ so that we can use the property of the delta function:
\begin{align}
\int_{\mathbb{R}}dxf\left(x\right)\braket{x|a+ib} & =\frac{1}{2\pi}\int_{\mathbb{R}}dpdxf\left(x\right)e^{ip\left(x-a-ib\right)}\nonumber \\
 & =\frac{1}{2\pi}\int_{\mathbb{R}}dpdxf\left(x+ib\right)e^{ip\left(x-a\right)}\nonumber \\
 & =f\left(a+ib\right).
\end{align}
At the first line we inserted the identity operator $\int_{\mathbb{R}}dp\kket p\bbra p$. This shows that we can calculate the integral \eqref{eq:CompDelta} regarding $\braket{x|a+ib}$ as the usual delta function as long as $f\left(x\right)$ satisfies the conditions above.

\section{Proof for quantum curves\label{sec:QCproof}}

In this appendix we prove \eqref{eq:QCgen}, where the left-hand side is defined in \eqref{eq:CurveDef}. We use the mathematical induction. The base case is $q=\overline{q}=0$ case and is easily proven. In the induction step, we prove that if the identity \eqref{eq:QCgen} holds for $\widehat{H}_{\bm{M}}^{\left(-\overline{q},+1,-q\right)}$, the identity also holds for $\widehat{H}_{\bm{M}'}^{\left(-\overline{q},+1,-\left(q+1\right)\right)}$ and $\widehat{H}_{\bm{M}''}^{\left(-\left(\overline{q}+1\right),+1,-q\right)}$, where
\begin{align}
\bm{M} & =\left(\overline{\mathcal{M}}_{\overline{q}},\ldots,\overline{\mathcal{M}}_{1},\mathcal{M}_{1},\ldots,\mathcal{M}_{q}\right),\nonumber \\
\bm{M}' & =\left(\overline{\mathcal{M}}_{\overline{q}},\ldots,\overline{\mathcal{M}}_{1},\mathcal{M}_{1}+M,\ldots,\mathcal{M}_{q}+M,M\right),\nonumber \\
\bm{M}'' & =\left(\overline{M},\overline{\mathcal{M}}_{\overline{q}}+\overline{M},\ldots,\overline{\mathcal{M}}_{1}+\overline{M},\mathcal{M}_{1},\ldots,\mathcal{M}_{q}\right).
\end{align}

First, we show that if the identity \eqref{eq:QCgen} holds for $\widehat{H}_{\bm{M}}^{\left(-\overline{q},+1,-q\right)}$, the identity also holds for $\widehat{H}_{\bm{M}'}^{\left(-\overline{q},+1,-\left(q+1\right)\right)}$. This assumption and \eqref{eq:QCabj} lead
\begin{align}
 & \left\{ \widehat{\rho}_{\bm{M}}^{\left(-\overline{q},+1,-q\right)}\left(\widehat{\rho}_{M}^{\left(+1,-1\right)}\right)^{\dagger}\right\} ^{-1}\nonumber \\
 & =\left(\widehat{H}_{M}^{\left(+1,-1\right)}\right)^{\dagger}\widehat{H}_{\bm{M}}^{\left(-\overline{q},+1,-q\right)}\nonumber \\
 & =\widehat{P}\widehat{\mathcal{Q}}_{M+k}\widehat{\mathcal{Q}}_{-\bm{M}_{\mathrm{R}}+k}\widehat{\mathcal{Q}}_{\bm{M}_{\mathrm{L}}}+\left(\widehat{\mathcal{Q}}_{-M+k}\widehat{\mathcal{Q}}_{-\bm{M}_{\mathrm{R}}+k}\widehat{\mathcal{Q}}_{\bm{M}_{\mathrm{L}}}+\widehat{\mathcal{Q}}_{M-k}\widehat{\mathcal{Q}}_{\bm{M}_{\mathrm{R}}-k}\widehat{\mathcal{Q}}_{-\bm{M}_{\mathrm{L}}}\right)\nonumber \\
 & \,\,\,+\widehat{P}^{-1}\widehat{\mathcal{Q}}_{-M-k}\widehat{\mathcal{Q}}_{\bm{M}_{\mathrm{R}}-k}\widehat{\mathcal{Q}}_{-\bm{M}_{\mathrm{L}}},
\end{align}
where we used \eqref{eq:QmPexchange}. We apply the similarity transformation generated by $\widehat{G}_{M}^{-1}$ defined in \eqref{eq:SimGen}, namely $\widehat{G}_{M}\left[\cdot\right]\widehat{G}_{M}^{-1}$ to the both side. The left-hand side becomes
\begin{align}
 & \left\{ i^{\mathcal{M}_{1}+M-\overline{\mathcal{M}}_{1}}\left(\prod_{\overline{a}}^{\overline{q}}\frac{\prod_{\overline{j}}^{\overline{M}_{\overline{a}}}2\sinh\frac{\widehat{q}-2\pi it_{\overline{M}_{\overline{a}},\overline{j}}}{2k}}{2\cosh\frac{\widehat{q}-i\pi M_{\overline{a}}}{2}}\right)\left(\prod_{a}^{q}\frac{1}{\prod_{t}^{M_{a}}2\cosh\frac{\widehat{q}-2\pi it_{M_{a},j}}{2k}}\right)\right.\nonumber \\
 & \,\,\,\times\frac{1}{\prod_{j}^{M}2\cosh\frac{\widehat{q}-2\pi it_{M,j}}{2k}}\frac{1}{2\cosh\frac{\widehat{p}}{2}}\left(\prod_{a}^{q}\frac{\prod_{j}^{M_{a}}2\sinh\frac{\widehat{q}-2\pi it_{M_{a},j}}{2k}}{2\cosh\frac{\widehat{q}+i\pi M_{a}}{2}}\right)\frac{\prod_{j}^{M}2\sinh\frac{\widehat{q}-2\pi it_{M,j}}{2k}}{2\cosh\frac{\widehat{q}+i\pi M}{2}}\nonumber \\
 & \left.\,\,\,\times\left(\prod_{\overline{a}}^{\overline{q}}\frac{1}{\prod_{\overline{j}}^{\overline{M}_{\overline{a}}}2\cosh\frac{\widehat{q}-2\pi it_{\overline{M}_{\overline{a}},\overline{j}}}{2k}}\right)\frac{1}{2\cosh\frac{\widehat{p}}{2}}\right\} ^{-1}\nonumber \\
 & =2\cosh\frac{\widehat{p}}{2}\left(\widehat{\rho}_{\bm{M}'}^{\left(-\overline{q},+1,-\left(q+1\right)\right)}\right)^{-1},
\end{align}
where we substituted \eqref{eq:rhoGen}, and the right-hand side becomes
\begin{align}
 & \widehat{\mathcal{Q}}_{-M-k}\widehat{P}\widehat{\mathcal{Q}}_{-\bm{M}_{\mathrm{R}}+k}\widehat{\mathcal{Q}}_{\bm{M}_{\mathrm{L}}}+\left(\widehat{\mathcal{Q}}_{-M+k}\widehat{\mathcal{Q}}_{-\bm{M}_{\mathrm{R}}+k}\widehat{\mathcal{Q}}_{\bm{M}_{\mathrm{L}}}+\widehat{\mathcal{Q}}_{M-k}\widehat{\mathcal{Q}}_{\bm{M}_{\mathrm{R}}-k}\widehat{\mathcal{Q}}_{-\bm{M}_{\mathrm{L}}}\right)\nonumber \\
 & \,\,\,+\widehat{\mathcal{Q}}_{M+k}\widehat{P}^{-1}\widehat{\mathcal{Q}}_{\bm{M}_{\mathrm{R}}-k}\widehat{\mathcal{Q}}_{-\bm{M}_{\mathrm{L}}}\nonumber \\
 & =\widehat{P}\widehat{\mathcal{Q}}_{-\bm{M}_{\mathrm{R}}'+k}\widehat{\mathcal{Q}}_{\bm{M}_{\mathrm{L}}'}+\left(\widehat{\mathcal{Q}}_{-\bm{M}_{\mathrm{R}}'+k}\widehat{\mathcal{Q}}_{\bm{M}_{\mathrm{L}}'}+\widehat{\mathcal{Q}}_{\bm{M}_{\mathrm{R}}'-k}\widehat{\mathcal{Q}}_{-\bm{M}_{\mathrm{L}}'}\right)\nonumber \\
 & \,\,\,+\widehat{P}^{-1}\widehat{\mathcal{Q}}_{\bm{M}_{\mathrm{R}}'-k}\widehat{\mathcal{Q}}_{-\bm{M}_{\mathrm{L}}'}\nonumber \\
 & =\left(\widehat{P}^{\frac{1}{2}}+\widehat{P}^{-\frac{1}{2}}\right)\left(\widehat{\mathcal{Q}}_{-\bm{M}_{\mathrm{R}}'}\widehat{P}^{\frac{1}{2}}\widehat{\mathcal{Q}}_{\bm{M}_{\mathrm{L}}'}+\widehat{\mathcal{Q}}_{\bm{M}_{\mathrm{R}}'}\widehat{P}^{-\frac{1}{2}}\widehat{\mathcal{Q}}_{-\bm{M}_{\mathrm{L}}'}\right),
\end{align}
where we used \eqref{eq:LargeSim}. Therefore, we obtain
\begin{equation}
\widehat{H}_{\bm{M}'}^{\left(-\overline{q},+1,-\left(q+1\right)\right)}=\widehat{\mathcal{Q}}_{-\bm{M}_{\mathrm{R}}'}\widehat{P}^{\frac{1}{2}}\widehat{\mathcal{Q}}_{\bm{M}_{\mathrm{L}}'}+\widehat{\mathcal{Q}}_{\bm{M}_{\mathrm{R}}'}\widehat{P}^{-\frac{1}{2}}\widehat{\mathcal{Q}}_{-\bm{M}_{\mathrm{L}}'}.
\end{equation}

Next, we show that if the identity \eqref{eq:QCgen} holds for $\widehat{H}_{\bm{M}}^{\left(-\overline{q},+1,-q\right)}$, the identity also holds for $\widehat{H}_{\bm{M}''}^{\left(-\left(\overline{q}+1\right),+1,-q\right)}$. The general flow of computation is similar to the above, and thus we omit the detail. We start with
\begin{align}
 & \left(\widehat{\rho}_{\overline{M}}^{\left(+1,-1\right)}\widehat{\rho}_{\bm{M}}^{\left(-\overline{q},+1,-q\right)}\right)^{-1}\nonumber \\
 & =\widehat{H}_{\bm{M}}^{\left(-\overline{q},+1,-q\right)}\widehat{H}_{\overline{M}}^{\left(+1,-1\right)}\nonumber \\
 & =\widehat{\mathcal{Q}}_{-\bm{M}_{\mathrm{R}}}\widehat{\mathcal{Q}}_{\bm{M}_{\mathrm{L}}-k}\widehat{\mathcal{Q}}_{-\overline{M}-k}\widehat{P}+\left(\widehat{\mathcal{Q}}_{-\bm{M}_{\mathrm{R}}}\widehat{\mathcal{Q}}_{\bm{M}_{\mathrm{L}}-k}\widehat{\mathcal{Q}}_{\overline{M}-k}+\widehat{\mathcal{Q}}_{\overline{M}-k}\widehat{\mathcal{Q}}_{\bm{M}_{\mathrm{R}}-k}\widehat{\mathcal{Q}}_{-\bm{M}_{\mathrm{L}}}\right)\nonumber \\
 & \,\,\,+\widehat{\mathcal{Q}}_{\bm{M}_{\mathrm{R}}}\widehat{\mathcal{Q}}_{-\bm{M}_{\mathrm{L}}+k}\widehat{\mathcal{Q}}_{\overline{M}-k}\widehat{P}^{-1}.
\end{align}
We apply the similarity transformation generated by $\widehat{G}_{M}$, namely $\widehat{G}_{M}^{-1}\left[\cdot\right]\widehat{G}_{M}$ to the both side. The left-hand side becomes
\begin{equation}
\left(\widehat{\rho}_{\bm{M}''}^{\left(-\left(\overline{q}+1\right),+1,-q\right)}\right)^{-1}2\cosh\frac{\widehat{p}}{2},
\end{equation}
and the right-hand side becomes
\begin{equation}
\left(\widehat{\mathcal{Q}}_{-\bm{M}_{\mathrm{R}}''}\widehat{P}^{\frac{1}{2}}\widehat{\mathcal{Q}}_{\bm{M}_{\mathrm{L}}''}+\widehat{\mathcal{Q}}_{\bm{M}_{\mathrm{R}}''}\widehat{P}^{-\frac{1}{2}}\widehat{\mathcal{Q}}_{-\bm{M}_{\mathrm{L}}''}\right)\left(\widehat{P}^{\frac{1}{2}}+\widehat{P}^{-\frac{1}{2}}\right).
\end{equation}
Therefore, we finally obtain
\begin{equation}
\widehat{H}_{\bm{M}''}^{\left(-\left(\overline{q}+1\right),+1,-q\right)}=\widehat{\mathcal{Q}}_{-\bm{M}_{\mathrm{R}}''}\widehat{P}^{\frac{1}{2}}\widehat{\mathcal{Q}}_{\bm{M}_{\mathrm{L}}''}+\widehat{\mathcal{Q}}_{\bm{M}_{\mathrm{R}}''}\widehat{P}^{-\frac{1}{2}}\widehat{\mathcal{Q}}_{-\bm{M}_{\mathrm{L}}''}.
\end{equation}
Thus \eqref{eq:QCgen} holds.

\sloppy

\printbibliography

\end{document}